\documentclass[fleqn,usenatbib]{mnras}

\usepackage{newtxtext,newtxmath}

\usepackage[T1]{fontenc}
\usepackage{ae,aecompl}

\usepackage{graphicx}	
\usepackage{amsmath}	
\usepackage{amssymb}	

\usepackage{rotating}
\usepackage[toc,page]{appendix}
\usepackage{verbatim}
\usepackage{subcaption}
\usepackage{float}
\usepackage{xcolor}

\usepackage{graphicx}	
\usepackage{amsmath}	
\usepackage{amssymb}	

\title[Atmospheric Retrieval of Brown Dwarfs]{Considerations for Atmospheric Retrieval of High-Precision Brown Dwarf Spectra}

\author[Piette \& Madhusudhan]{
Anjali A. A. Piette,$^{1}$\thanks{E-mail: ap763@cam.ac.uk}
Nikku Madhusudhan,$^{1}$\thanks{E-mail: nmadhu@ast.cam.ac.uk}
\\
$^{1}$Institute of Astronomy, University of Cambridge, Madingley Road, Cambridge, CB3 0HA, UK\\
}

\date{Accepted XXX. Received YYY; in original form ZZZ}

\pubyear{2019}

\begin{document}
\label{firstpage}
\pagerange{\pageref{firstpage}--\pageref{lastpage}}
\maketitle

\begin{abstract}
Isolated brown dwarfs provide remarkable laboratories for understanding atmospheric physics in the low-irradiation regime, and can be observed more precisely than exoplanets. As such, they provide a glimpse into the future of high-SNR observations of exoplanets. In this work, we investigate several new considerations that are important for atmospheric retrievals of high-quality thermal emission spectra of sub-stellar objects. We pursue this using an adaptation of the HyDRA atmospheric retrieval code. We propose a parametric pressure-temperature (P-T) profile for brown dwarfs consisting of multiple atmospheric layers, parameterised by the temperature change across each layer. This model allows the steep temperature gradient of brown dwarf atmospheres to be accurately retrieved while avoiding commonly-encountered numerical artefacts. The P-T model is especially flexible in the photosphere, which can reach a few tens of bar for T-dwarfs. We demonstrate an approach to include model uncertainties in the retrieval, focusing on uncertainties introduced by finite spectral and vertical resolution in the atmospheric model used for retrieval ($\sim$8\% in the present case). We validate our retrieval framework by applying it to a simulated data set and then apply it to the HST/WFC3 spectrum of the T-dwarf 2MASS~J2339+1352. We retrieve sub-solar abundances of H$_2$O and CH$_4$ in the object at $\sim$0.1~dex precision. Additionally, we constrain the temperature structure to within $\sim$100~K in the photosphere. Our results demonstrate the promise of high-SNR spectra to provide high-precision abundance estimates of sub-stellar objects.

\end{abstract}




\section{Introduction}

The field of exoplanet and brown dwarf atmospheres is undergoing rapid progress, with increasingly detailed characterisation made possible by state-of-the-art spectroscopic observations and sophisticated retrieval methods. Thanks to the high precision and wide spectral coverage of these observations, they allow us to probe a range of altitudes and determine the chemical properties and physical structures of these objects \citep[e.g.][]{Basri2000,Kirkpatrick2005,Apai2013,Buenzli2014,Helling2014,Sing2016,Burningham2017,Line2017,Madhusudhan2018}.
Directly-imaged brown dwarfs can provide a glimpse into the future of exoplanet observations as their lack of stellar companion and prominent spectral features make them ideal for very high-precision observations and characterisation, while their atmospheres bear many similarities with those of giant planets in the low-irradiation regime \citep[e.g.][]{Basri2006,Burgasser2009,Marley2015}.

Most of the exoplanet atmospheres characterised to date belong to hot Jupiters, whose large scale-heights and high temperatures ($\sim$800 - 3000K) make them conducive to atmospheric observations. Both space-based and ground-based observations have allowed a number of chemical species to be observed in hot Jupiters and constraints have also been placed on their physical structures \citep[e.g.][]{Brogi2012,Deming2013,Evans2017}. Chemical constraints have also been placed for several directly-imaged giant planets \citep[e.g.][]{Konopacky2013,Snellen2014,Macintosh2015,Todorov2016,Lavie2017,GravityCollaboration2020}. Such chemical constraints have the potential to provide insights into the formation conditions of exoplanets \citep[e.g.][]{Madhusudhan2016}.

Isolated brown dwarfs bear many similarities to directly-imaged giant planets, which have wide orbital separations and so receive negligible stellar irradiation \citep[e.g.][]{Burgasser2009,Marley2015}. Both classes of objects have H$_2$-rich atmospheres, and with no irradiation their spectra are dominated by internal heat. Consequently, the physical processes in both classes of object may be expected to be similar for comparable effective temperatures, making brown dwarfs exquisite analogs to investigate the physical characteristics of exoplanets at large orbital separations.

Isolated brown dwarfs are also excellent probes of unknown sub-stellar formation mechanisms, as they lie in the transitional domain between planetary and stellar masses where such mechanisms are still in debate \citep[e.g.][]{Whitworth2007,Luhman2012,Chabrier2014}. Determining their compositions may contribute to constraining such formation mechanisms by finding (dis)similarities with planetary and stellar properties.

Similarly to exoplanets, brown dwarf spectra have traditionally been analysed by identifying specific molecular features or fitting them with grids of equilibrium models \citep[e.g.][]{Mohanty2004a,Mohanty2004b}. Based on the presence of various combinations of species in different objects, this has led to the classification of brown dwarfs into the spectral types L, T and Y, analogous to those of stars \citep{Kirkpatrick2005, Cushing2011}. Although the L-T classification is based only on spectral features, estimates of the effective temperatures of L and T dwarfs \citep{Vrba2004,Dahn2002, Golimowski2004,Kirkpatrick2005} have shown that the spectral sequence follows a sequence in temperature, where early L dwarfs are hottest and late T dwarfs are coolest. Furthermore, L dwarfs show signs of clouds and dust which are not typically seen in T-dwarfs \citep[e.g.][]{Marley2002,Tsuji2002}. Cloud formation and sedimentation, as well as dynamical processes and disequilibrium chemistry, have been explored through both self-consistent models and observations of brown dwarf spectra \citep[e.g.][]{Saumon2000,Burrows2006, Cushing2008,Apai2013,Marley2015}.

More recently - as with exoplanet atmospheres - atmospheric retrievals have begun to be used with brown dwarf spectra \citep[e.g.][]{Line2014b, Line2015, Line2017, Burningham2017,Zalesky2019,Kitzmann2020}. Since brown dwarfs closely resemble giant planets, the retrieval methods used are similar, though with some notable differences. For example, unlike exoplanets, the gravities and radii of brown dwarfs are not typically known and these quantities are included as free parameters in the retrieval. Furthermore, since isolated sub-stellar objects are not expected to exhibit thermal inversions or deep isotherms, their $P$-$T$ profiles can be parameterised differently to those of irradiated exoplanets \citep[e.g.][]{Line2015, Burningham2017,GravityCollaboration2020,Kitzmann2020}. Retrieval methods also differ between L and T dwarfs, as L dwarfs are known to have clouds, which must therefore be included in the retrieval analyses of their spectra \citep{Burningham2017}. In contrast, \citet{Line2015} show that the inclusion of clouds in their retrievals of T-dwarf spectra does not affect their results .

One of the principal differences between brown dwarf retrieval approaches in the literature is the treatment of the pressure-temperature ($P$-$T$) profile \citep[e.g.][]{Burningham2017,Line2017,Kitzmann2020}. For example, \citet{Line2015,Line2017} use 15 temperature parameters, corresponding to fixed pressures, and interpolate between them in order to compute radiative transfer at a higher resolution. Since these temperature parameters are degenerate, they also penalise the second derivative of the temperature profile in order to prevent unphysical oscillations in temperature. \citet{Burningham2017}, however, find that this method - originally used with T-dwarfs - presents challenges with cloudy L-dwarfs as the spectral contrast of their data is not able to shift the $P$-$T$ profile away from the linear fit that is preferred by the second derivative minimisation. Instead, \citet{Burningham2017} use the parametric $P$-$T$ model from \citet{Madhusudhan2009}. \citet{Kitzmann2020} consider two further approaches to model the $P$-$T$ profile: a piecewise polynomial and an approximate solution to radiative transfer assuming radiative equilibrium in a gray atmosphere, both chosen to avoid unphysical oscillations.

In this work, we present various considerations that can affect the accuracy of brown dwarf retrievals, focusing especially on T-dwarfs. Motivated by the high-quality spectra available for brown dwarfs, we seek ways to maximise the precision and accuracy of the results obtained. We begin by introducing a way of considering uncertainties in the retrieval, in particular taking into account the uncertainties in the model. We find that it can be important to explicitly factor in model uncertainty in order to accurately retrieve the properties of an atmosphere from a high signal-to-noise-ratio (SNR) spectrum. We also introduce a new multi-layer $P$-$T$ parametrisation in which the average slope of each layer is a parameter and the temperature nodes between the layers are interpolated such that the first derivative of the $P$-$T$ profile is continuous throughout the atmosphere. The priors on the slope parameters, as well as the interpolation method, are chosen to prevent numerical artefacts. We implement this $P$-$T$ model in the HyDRA retrieval code \citep{Gandhi2018}. We then test our approach on simulated data and apply it to a high-SNR spectrum of the T-dwarf 2MASS~J2339+1352.

We begin by giving a general outline of the retrieval framework we use in section \ref{sec:model}. Our developments in considering uncertainties and modeling the $P$-$T$ profile are then described in sections \ref{sec:uncertainties} and \ref{sec:PT}, respectively. We test the method on synthetic data in section \ref{sec:self-consistency} and apply it to a real T-dwarf spectrum in section \ref{sec:results}. Our conclusions and discussion are presented in section \ref{sec:discussion}.

\section{Atmospheric Model}
\label{sec:model}

In this work, we use the retrieval method of \citet{Gandhi2018} (HyDRA), with some modifications. This method involves a parametric atmospheric forward model coupled to PyMultiNest \citep{Feroz2009,Buchner2014}, a nested sampling Bayesian parameter estimation algorithm \citep{Skilling2006}. For a given $P$-$T$ profile and chemical abundances, the forward model calculates the corresponding spectrum. HyDRA has thus far been used to retrieve the thermal emission spectra of transiting exoplanets. Here, we adapt the retrieval method to apply it to high-SNR brown dwarf spectra. In particular, HyDRA uses a parametric $P$-$T$ profile (similar to that of \citet{Madhusudhan2009}) which allows for features that can be present in irradiated atmospheres, such as deep isotherms and thermal inversions. In this work, we develop a new parametric $P$-$T$ profile as the spectra we consider have a much higher SNR than typical exoplanet spectra. As a result, the data is better able to constrain the $P$-$T$ profile and a more flexible $P$-$T$ model is desirable to fully capture the information in the spectrum. In addition to a new $P$-$T$ model, we also adapt the retrieval method to include sources of model uncertainty, which can become significant when observational uncertainty is very small.

Another important difference between transiting planets and directly imaged brown dwarfs is that the radius and mass of transiting planets can be obtained through primary transit and radial velocity observations, and the distance to the host star (and therefore the planet) is typically known. In contrast, these quantities are unknown for isolated brown dwarfs and, hence, need to be included as free parameters in the model. We include them as the radius-distance ratio ($R$/$d$) and log gravity (log($g$)) as these quantities are independent in the calculation of the spectrum (note that radius and distance only affect the spectrum through the quantity $R$/$d$ so are perfectly degenerate, as shown in equation \ref{eq:Fnu}).

Since giant planets and brown dwarfs both have H$_2$-rich atmospheres, the chemical species we include in the forward model are largely similar to those in exoplanet studies. To decide which species to include, we turn to previous analyses of T-dwarf spectra, which have shown that H$_2$O, CH$_4$ and collision-induced absorption (CIA) from H$_2$-H$_2$ and H$_2$-He interactions dominate their spectra \citep{Kirkpatrick2005}. Na and K are also known to be present in the near-infrared spectra of T-dwarfs, especially towards shorter wavelengths in the wings of their strong optical lines. Although CO is not expected to occur in equilibrium at the temperatures in T-dwarfs, it has been observed in some cases including the first discovered T dwarf, GJ 229b \citep{Noll1997, Kirkpatrick2005}, and is likely indicative of atmospheric vertical mixing \citep{Moses2011}. NH$_3$ is also expected to appear in T-type spectra, although not as strongly as H$_2$O and CH$_4$ \citep{Roellig2004}. For completeness, we also retrieve the abundances of CO$_2$ and HCN which could potentially be significant in H$_2$-rich atmospheres. The cross sections we use in our models, apart from those of Na and K, are calculated as in \citet{Gandhi2017} from the HITEMP, HITRAN and ExoMol line list databases (H$_2$O, CO and CO$_2$: \citet{Rothman2010}, CH$_4$: \citet{Yurchenko2013,Yurchenko2014a}, NH$_3$: \citet{Yurchenko2011}, HCN: \citet{Harris2006,Barber2014}, CIA: \citet{Richard2012}). The broadening we use for Na and K is described in detail in section \ref{sec:model uncertainties}. Note that, since clouds are not typically detected in T-dwarfs, we do not include them in the model (e.g. \citealt{Tsuji2002,Kirkpatrick2005,Line2015}, but see also \citealt{Morley2012,Buenzli2014}).

The forward model spectrum is generated from the $P$-$T$ parameters, chemical abundances and bulk parameters by solving radiative transfer, hydrostatic equilibrium and the ideal gas law successively across thin layers of the atmosphere \citep{Gandhi2018}:
\begin{align}
\mu \frac{\textrm{d}I_{\nu}}{\textrm{d}\tau} &= I_{\nu}-B_{\nu} \label{eq:RT}\\
\textrm{d}\tau_{\nu} &= \sum_i \sigma_i n_i \textrm{d}z \label{eq:dtau}\\
\frac{\textrm{d}P}{\textrm{d}z} &= -\rho g \label{eq:hydrostatic_eq}\\
P &=n k_B T \label{eq:ideal_gas}
\end{align}
where $P$, $T$, $\rho$, $n$, $\tau_{\nu}$ and dz are the pressure, temperature, gas mass density, gas number density, optical depth and thickness of the atmospheric layer, respectively. g is the gravitational field at the radius of the layer, $I_{\nu}$ is specific intensity, $B_{\nu}(T)$ is the Planck function and $\mu$=cos($\theta$) where $\theta$ is the angle of a ray relative to the normal. In equation \ref{eq:dtau}, the sum is over all species in the atmosphere and  $\sigma_i$ and $n_i$ are the cross section and number density of the $i^{th}$ species, respectively. In the retrieval, the abundances of the chemical species are characterised by their mixing ratios, $X_i = n_i/n$. We nominally use 100 thin atmospheric layers for the radiative transfer calculation, but investigate the effect of this number on the accuracy of the spectrum in section \ref{sec:model uncertainties}.

At the top of the atmosphere, the emergent flux, F$_{\nu,\textrm{em}}$, is
\begin{equation*}
F_{\nu,\textrm{em}} = 2 \pi \int_{0}^{1} I_{\nu,\textrm{em}}(\mu) \, \mu  \, \textrm{d} \mu
\end{equation*}
where $I_{\nu,\textrm{em}}$ is the emergent spectral intensity. The flux at the observer is then
\begin{equation}
\label{eq:Fnu}
F_{\nu} = \frac{R^2}{d^2}  \, F_{\nu,\textrm{em}}.
\end{equation}
This model spectrum is convolved with HST's PSF and then binned into simulated data points at the resolution of HST/WFC3 data.

The parameters listed above are estimated for a given observed spectrum using the Nested Sampling algorithm \citep{Skilling2006}, implemented using PyMultiNest \citep{Buchner2014}. The algorithm estimates the posterior probability distributions of each parameter. These posteriors are given by Bayes' Theorem:
\begin{equation*}
P(\theta|d,M) = \frac{\mathcal{L}(d|\theta,M) \,\, \pi(\theta)}{P(d|M)},
\end{equation*}
where $\theta$ is the vector of model parameters, $d$ is the data, $M$ is the model, $P(\theta|d,M)$ is the posterior distribution, $\mathcal{L}(d|\theta,M)$ is the likelihood function, $\pi(\theta)$ are the priors on the parameters and $P(d|M)$ is the Bayesian evidence. The priors used for each parameter are shown in table \ref{tab:priors}. One of the advantages of nested sampling over other Bayesian parameter estimation methods is that the Bayesian evidence of the model is evaluated to high accuracy, allowing different models to be compared and the detection significances of chemical species to be evaluated statistically. In this work, we assume a Gaussian form for the likelihood function:

\begin{equation}
\label{eq:likelihood}
\mathcal{L}(d|\theta) = \sum_{i}\frac{1}{\sqrt{2 \pi \sigma_i^2}} \,\,  \mathrm{exp}\left[{-\frac{\left(\textrm{model}_i-\textrm{data}_i\right)^2}{2\sigma_i^2}}\right],
\end{equation}
where the summation is over all data points and the uncertainties, $\sigma_i$, will be discussed in section \ref{sec:uncertainties}.

\begin{table}
\begin{centering}
\begin{tabular}{c|c|c}
Parameter & Prior & Range \\
\hline
$X_i$ & log-uniform & $10^{-10} - 10^{-2}$\\
$R/d \,\, (R_\mathrm{J}/\mathrm{pc})$ & uniform & $10^{-3}$ - 1 \\
log($g$/cms$^{-2}$) & uniform & 2 - 7\\
$x_{tol}$ & uniform & 8\% - 100\% \\
$T_{3.2\mathrm{b}}$ (K) & uniform & 300 - 4000 \\
$\Delta T_{100-32\mathrm{b}}$ (K) & uniform & 0 - 2500 \\
$\Delta T_{32-10\mathrm{b}}$ (K) & uniform & 0 - 2000 \\
$\Delta T_{10-3.2\mathrm{b}}$ (K) & uniform & 0 - 1500 \\
$\Delta T_{3.2-1\mathrm{b}}$ - $\Delta T_{10-1\mathrm{mb}}$ (K) & uniform & 0 - 1000 \\
\end{tabular}
\caption{\label{tab:priors} The priors used for each model parameter. All $\Delta T_{i}$ parameters below pressures of 3.2 bar have the same prior range. }
\end{centering}
\end{table}

\begin{figure}
\centering
    	\includegraphics[width=0.5\textwidth]{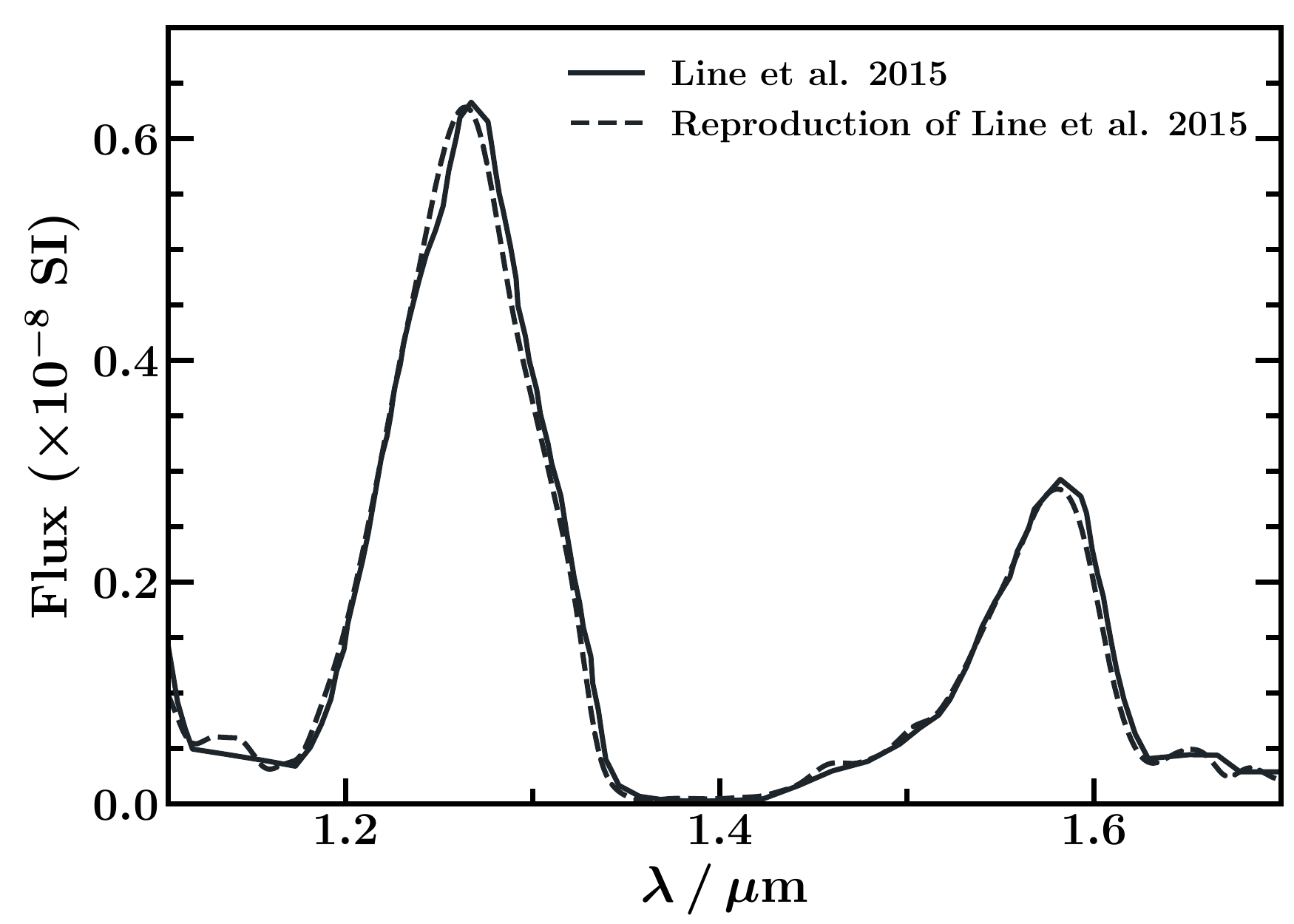}
\caption{\label{fig:Linecomp} Model spectra for validation. The retrieved spectrum of Gl~570D by \citet{Line2015} is shown by the solid line, and our reproduction of this spectrum using the same $P$-$T$ profile and atmospheric parameters is shown by the dashed line. The two spectra are consistent with each other. Flux axis is shown in SI units (Wm$^{-2}$m$^{-1}$).}
\end{figure}

In order to test this retrieval method - and various $P$-$T$ models within it - we create a simulated spectrum with known input parameters (black spectrum in right panel of figure \ref{fig:sim_data_depth_origin}). A successful method should then be able to accurately retrieve these inputs. Throughout this work, we base our simulated spectrum on the well-studied T-dwarf Gl~570D \citep[e.g.][]{Burgasser2000, Burgasser2003, Burgasser2006, Geballe2001, Geballe2009, Leggett2002, Cushing2006, Patten2006, Saumon2006, Hubeny2007, Line2014b, Line2015}. For the input $P$-$T$ profile, we use the equilibrium $P$-$T$ profile shown in figure 3 of \citet{Line2015} (based on \citet{Saumon2008}, shown here in the left panel of figure \ref{fig:sim_data_depth_origin}). The abundances for H$_2$O, CH$_4$, NH$_3$, CO, CO$_2$, Na and K  as well as the radius-distance ratio and gravity are those retrieved by \citet{Line2015}, and are listed in table \ref{tab:sim_data}. In order to test our forward model, we recreate the median retrieved spectrum from \citet{Line2015} using their median retrieved $P$-$T$ profile (different from the equilibrium $P$-$T$ profile that we use for our simulated data) and the parameters listed in table \ref{tab:sim_data}. The spectrum we generate with these inputs is consistent with that of \citet{Line2015}, validating our forward model (figure \ref{fig:Linecomp}).


\begin{table}
\begin{centering}
\begin{tabular}{c|c|c|c}
Parameter & Input value & Retrieved value\\[3pt]
 \hline
 $\textrm{log}(X_{H_2O})$ & -3.45 &$-3.4 \pm 0.1$  \\[7pt]
 $\textrm{log}(X_{CH_4})$ & -3.40 &  $-3.3 \pm 0.1$ \\[7pt]
 $\textrm{log}(X_{NH_3})$ & -4.64 &  $-4.6 \pm 0.1$\\[7pt]
 $\textrm{log}(X_{CO})$ & -7.53 &  $ - $\\[7pt]
 $X_{HCN}$ & 0.00 &  $ - $ \\[7pt]
 $\textrm{log}(X_{CO_2})$ & -7.76 &  $ - $  \\[7pt]
 $\textrm{log}X_{Na}$ & -5.50 &  $ - $ \\[7pt]
 $\textrm{log}X_{K}$ & -6.69 &  $6.8^{+0.3}_{-1.1}$ \\[7pt]
 $R/d\,\, (R_\mathrm{J}/\mathrm{pc})$ & 0.1952 & $0.20 \pm 0.01$ \\[7pt]
 log($g$/cms$^{-2}$) & 4.76 & $4.8 \pm 0.2$ \\[7pt]
 $x_{tol}$ & - &  $8.05^{+0.07}_{-0.03} $ \% \\[7pt]
\end{tabular}
\caption{\label{tab:sim_data} Atmospheric parameters used to simulate the HST/WFC3 spectrum of Gl~570D, from the retrieval of \citet{Line2015}. Also shown are the parameter values we obtain from the retrieval of this simulated data, which are consistent with the true input values (section \ref{sec:self-consistency}). CO, HCN, CO$_2$ and Na are not constrained by the retrieval.}
\end{centering}
\end{table}

\section{Treatment of Model Uncertainties}

\label{sec:uncertainties}
Retrievals performed on exoplanet and brown dwarf spectra to date have had sufficiently low SNR that model uncertainties may not have been significant. However, for high-precision spectra, it can be important to consider model uncertainties in addition to observational uncertainties. The data considered in this work, obtained by the Hubble Space Telescope's Wide-Field Camera 3 (HST/WFC3), has an extremely high SNR at only a $\sim$0.1\% noise level. The uncertainties in models could therefore add a potentially significant contribution to the $\sigma_i$ term in the likelihood function (equation \ref{eq:likelihood}). We discuss sources of model uncertainty in section \ref{sec:model uncertainties}, and quantify the magnitude of uncertainty due to resolution effects. In section \ref{sec:sim_data}, we describe how simulated data are generated such that the effects of these uncertainties on a retrieval can be tested (section \ref{sec:self-consistency}). In section \ref{sec:tol}, we discuss how quantified model uncertainties can be incorporated in a retrieval, along with a `tolerance' parameter which captures unknown sources of model and/or observational uncertainty.

\begin{figure}
\centering
    	\includegraphics[width=0.5\textwidth]{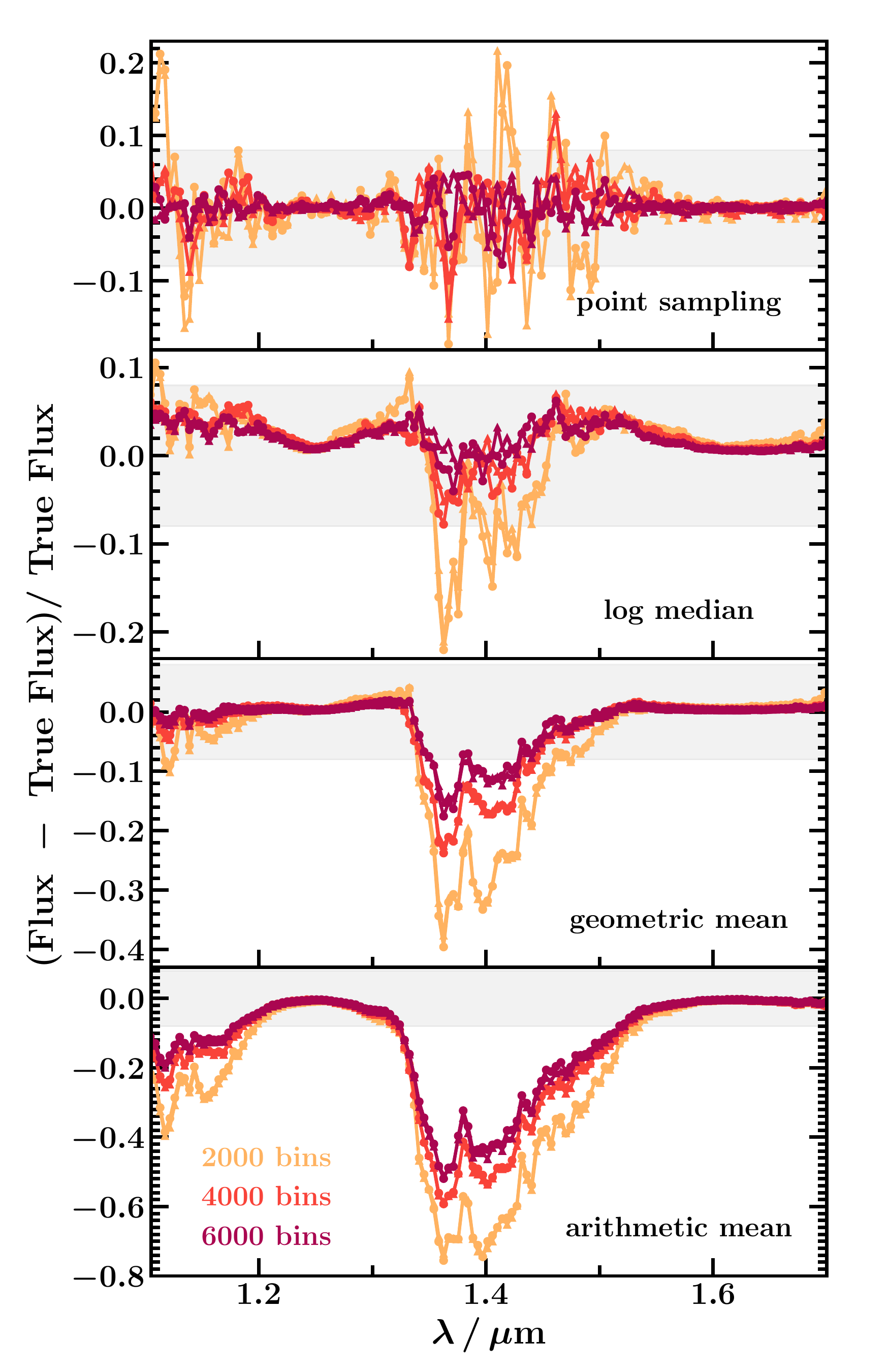}
\caption{\label{fig:rescomp} Fractional error in the spectrum for different cross section sampling methods. Orange, red and maroon lines and markers correspond to different spectral resolutions, i.e. 2000, 4000 and 6000 linearly-spaced wavelength bins in the range 1.1-1.7~$\mu$m, respectively. Whether the cross sections being sampled have a native resolution of 0.01cm$^{-1}$ (triangles) or 0.1cm$^{-1}$ (circles) does not have a significant effect on the spectral error. The region of $\leq$8\% error is shaded in grey.}
\end{figure}

\begin{figure}
\centering
    	\includegraphics[width=0.45\textwidth]{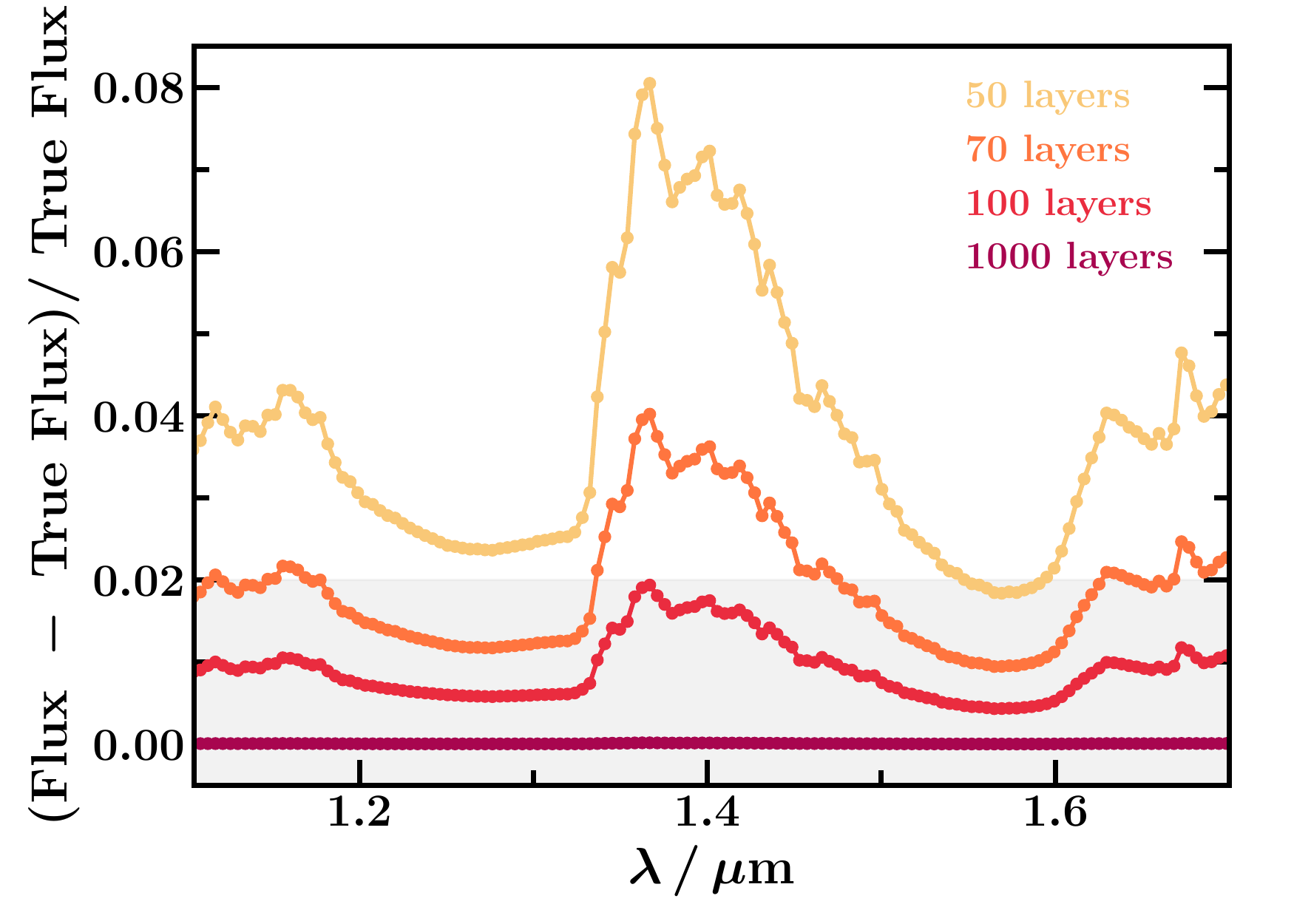}
\caption{\label{fig:NDcomp} Fractional error in the spectrum for different numbers of atmospheric layers in the model (between 100 and $10^{-3}$ bar). The region of $\leq$2\% error is shaded in grey.}
\end{figure}
\begin{figure*}
\centering
    	\includegraphics[width=\textwidth]{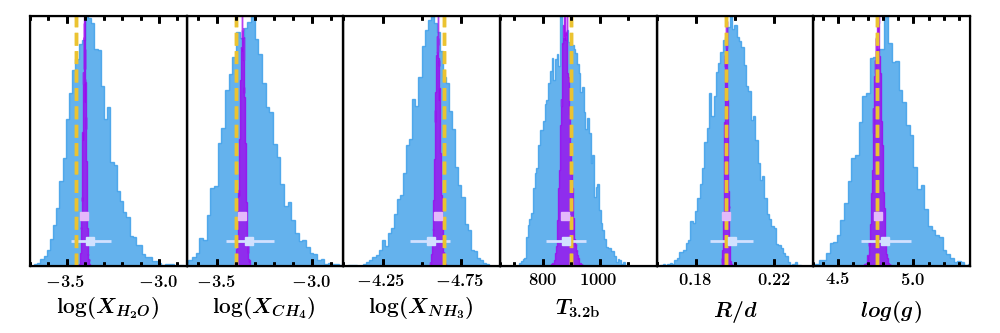}
\caption{\label{fig:tolcomp} Posterior probability distributions obtained with two different tolerance parameters. Squares and error bars show the median retrieved parameter values and 68\% confidence intervals, respectively, and the true input values are shown by yellow dashed lines. Purple histograms and light purple squares (top) show the retrieval output when the tolerance parameter takes the form in equation \ref{eq:ctol}, with a prior which allows a negligibly small value of the tolerance to be retrieved (note that the error bars are smaller than the symbol). In this case, the uncertainty in the retrieved parameters is noticeably underestimated. Blue histograms and light blue squares and error bars (bottom) correspond to the tolerance parameter described in equation \ref{eq:mtol}, with a prior which explicity accounts for known model uncertainties. This form of the tolerance parameter successfully retrieves the atmospheric parameters to within one standard deviation.}
\end{figure*}
\subsection{Uncertainties in the model spectra}
\label{sec:model uncertainties}

In this work, we focus on quantifying two important sources of model uncertainty: spectral and vertical resolution. In particular, we explore various methods and spectral resolutions for sampling molecular cross sections, as well as the number of atmospheric layers used to calculate the radiative transfer (i.e. vertical resolution). In this section, we assess the magnitude of these uncertainties. As a compromise between precision and computational runtime, we choose to use spectral and vertical resolutions which result in a combined uncertainty of $\sim8\%$ in the spectrum. We also consider uncertainties due to the broadening of Na and K lines and briefly discuss other potential sources of model error which could be addressed in future work.

\textbf{Spectral resolution: }Since the calculation of each model in the retrieval must be very fast, the resolution to which a model spectrum can be calculated (before binning) is limited. The molecular cross sections are typically computed at very high resolution with a grid spacing of 0.01-0.1cm$^{-1}$. The cross sections of each species must be sampled at lower resolution, which affects the accuracy of the spectrum. This can be done in several ways, e.g. sampling individual points, log-averaging the cross sections over wavelength bins or using the correlated-k method \citep[e.g.][]{Irwin2008,Amundsen2017} . Here we compare the accuracies of four sampling methods:
\begin{enumerate}
    \item point-sampling: values of the cross sections are taken at the chosen spectral resolution
    \item geometric mean : the cross sections are divided into bins at the chosen spectral resolution, and the representative cross sections in each bin are log-averaged from the native resolution. The averaged cross section is associated with the central wavelength of the bin.
    \item taking the arithmetic mean: similar to (ii) , but the arithmetic mean is taken rather than log-averaging.
    \item taking the log median : similar to (ii)  again, but the median log cross section of the bin is used rather than the log average.
\end{enumerate}

We also wish to test how the native resolution of the cross sections affects the accuracy of the spectrum, i.e. is sampling from a 0.1cm$^{-1}$ resolution cross section less accurate than sampling from a 0.01cm$^{-1}$ resolution cross section? Since the model spectra are ultimately binned to the resolution of the data (R$\sim$100), spectral information is lost and using very high-resolution cross sections is likely unnecessary. In our models, we use molecular cross sections for the various species at a native spacing of 0.1cm$^{-1}$. Here, we also test models with an even higher resolution (spacing of 0.01cm$^{-1}$) to evaluate its effect on spectral accuracy. Since the cross section profile of CIA from H$_2$ and He is very smooth, we use a resolution of 1cm$^{-1}$  for CIA opacity  throughout this work; a higher resolution is not needed as no sharp absorption features are revealed by doing this. In order to test the effect of the native resolution on the spectrum, we use a model atmosphere including only H$_2$, He and H$_2$O and compare the 0.01cm$^{-1}$ and 0.1cm$^{-1}$ native resolutions of the H$_2$O cross sections. We generate the spectra assuming a H$_2$O mixing ratio of $10^{-3.45}$ and using the $P$-$T$ profile shown in the left panel of figure \ref{fig:sim_data_depth_origin}.

In figure \ref{fig:rescomp}, we show the effect of spectral resolution and sampling method on the accuracy of spectra. We consider three sampling resolutions: 2000, 4000 and 6000 linearly-spaced wavelength points in the range 1.1-1.7$\mu$m (i.e. R$\sim$ 5000, 10000 and 15000 in the WFC3 band). We investigate the four sampling methods mentioned previously, which are performed on cross sections of H$_2$O for which the native resolution is either 0.1cm$^{-1}$ (shown by squares) or 0.01cm$^{-1}$ (shown by triangles). The accuracy of these spectra is evaluated against a spectrum generated at native resolution using the higher-resolution (0.01cm$^{-1}$) H$_2$O cross sections.

We confirm that, as expected, the native resolution of the cross sections which are sampled (between 0.01cm$^{-1}$ - 0.1cm$^{-1}$) makes very little difference to the binned spectrum. The effects of the four sampling methods are, however, quite different. The geometric and arithmetic mean methods result in very large inaccuracies in the spectrum in some wavelength ranges, so we choose not to use these methods. Point-sampling and taking the log median  introduce comparable magnitudes of uncertainty to the spectrum, but taking the log median  performs slightly better; using 4000 wavelength points results in up to >10\% uncertainty with point-sampling, but $\leq \sim$8\% with using  the log median . Using the log median  sampling method therefore results in a better accuracy pay-off for a given spectral resolution of the model.

We therefore choose to use the median sampling method in our retrieval framework. The number of wavelength bins used to compute the model spectra is then a compromise between computation time and accuracy. We choose to use 4000 bins as the $\leq \sim$8\% uncertainty it introduces still allows very good constraints to be placed on the atmospheric model parameters, while preserving a reasonable computation time.

\textbf{Vertical resolution: }
As described in section \ref{sec:model}, equations \ref{eq:RT}-\ref{eq:ideal_gas} are solved iteratively through each discrete layer in the model atmosphere between a maximum and minimum pressure. These pressure limits should be sufficiently generous that the spectrum is not affected by them, but narrow enough that not too many layers are needed to cover the pressure range. The minimum pressure, $P_{min}$, is dictated by the pressure at which the optical depth of the atmosphere is low enough not to affect the spectrum significantly. From figure \ref{fig:sim_data_depth_origin}, we see that using $P_{min}=10^{-3}$ should generously allow the photosphere to be probed. Since the spectrum is not affected by the deep regions of the atmosphere where the optical depth is very high, $P_{max}$ should lie deeper than the deepest pressure which affects the spectrum. Figure \ref{fig:sim_data_depth_origin} shows that contributions to the spectrum become small below $\sim$70 bar, making 100 bar a suitable choice which amply captures the photosphere.

Given this pressure range, increasing the number of atmospheric layers increases the accuracy of the model spectrum, akin to the improvement in accuracy of a numerical integrator by decreasing its step size. However, since the computation time of one model spectrum scales linearly with the number of atmospheric layers, the optimal number of layers should be the minimum number needed for the contribution to the model uncertainty to be relatively insignificant. Figure \ref{fig:NDcomp} shows that using 100 layers between 100 bar and $10^{-3}$ bar only contributes $\sim$2\% uncertainty to the model, and this also allows for a manageable computation time. We therefore choose to calculate model spectra in the retrieval using 100 layers between 100 bar and $10^{-3}$ bar.

Our chosen spectral and vertical resolutions in the current models therefore result in a combined $\sim$8\% uncertainty when added in quadrature. Given that the majority of the error is from the spectral resolution, this model uncertainty could be reduced if we were to consider higher spectral resolution, albeit at a greater computational cost. In what follows, we consider that the uncertainty in our present model has a lower limit of $\sim$8\%. In section \ref{sec:tol}, we describe how this lower limit is included in our retrievals.

\textbf{Na and K cross sections: }

Na and K both have strong features in the optical (at $\sim$0.6$\mu$m and $\sim$0.8$\mu$m, respectively) with broad wings which extend into the near-infrared, as well as smaller lines in the optical and infrared. In particular, the K lines at $\sim$1.25 $\mu$m produce a clear spectral feature present in the spectra of several T-dwarfs, including that of 2MASS J2339+1352 \citep{Buenzli2014} which we discuss in section \ref{sec:results}. In the past, various treatments have been used for the broadening of both the strong optical and weaker infrared lines \citep{Burrows2003,Allard2007}. We use the recent line profiles of \citet{Allard2016} and \citet{Allard2019} for the strong lines. However, for the smaller lines, there is some uncertainty as to which broadening profile is most appropriate.

In this work, we attempt to address the uncertainty in this broadening by using a simple parameterisation for the cross sections of the strongest K lines in the spectral range of interest. We apply the broadening profiles of \citet{Allard2016}, scaled by line strength \citep{NIST_ASD}, to the K lines at 1.169, 1.177 ,1.178 ,1.244 and 1.253$\mu$m, and modulate their cross sections with a multiplicative factor which is a free parameter in the retrieval. We use two multiplicative factors: one ($m_\mathrm{K1}$) for the lines at 1.244 and 1.253 $\mu$m and another ($m_\mathrm{K2}$) for the group of lines at 1.169, 1.177 and 1.178 $\mu$m. To avoid degeneracies between these parameters and the retrieved abundance of K ($X_\mathrm{K}$), we choose to retrieve $X_\mathrm{K1}=m_\mathrm{K1}\times X_\mathrm{K}$ and $X_\mathrm{K2}=m_\mathrm{K2}\times X_\mathrm{K}$. As well as the opacity due to these K lines, we also include the wings of the broadened optical lines at $\sim$0.6$\mu$m and $\sim$0.8$\mu$m from Na and K, respectively, which we parameterise according to the abundances of Na and K ($X_\mathrm{Na}$ and $X_\mathrm{K}$, respectively).

We apply this parameterisation in the retrieval of the HST/WFC3 spectrum of 2MASS J2339+1352 (section \ref{sec:results}). In the spectral range we consider in this work (1.1-1.7 $\mu$m), the wing of the Na optical line is fairly weak so we do not expect a strong constraint on the abundance of Na. However, since the K lines at $\sim$1.25 $\mu$m correspond to a noticeable feature in the spectrum of 2MASS J2339+1352, we expect that $X_\mathrm{K1}$ will be strongly constrained. Note that without a detection of the K wing (i.e. a constraint on $X_\mathrm{K}$), the retrieved value of $X_\mathrm{K1}$ does not result in an estimate of $m_\mathrm{K1}$; instead, this parameter simply allows the K feature at $\sim$1.25 $\mu$ to be fitted. More data at shorter wavelengths, where the wings of the  Na and K optical lines are stronger, would allow better constraints on the absolute abundances of these species.

\textbf{Other sources of uncertainty: }
Although we only quantify two sources of model uncertainty in this work, we stress that other sources of error can also play an important role. This is especially true if uncertainty due to prominent effects (e.g. spectral resolution) is substantially reduced (e.g. by significantly increasing the spectral resolution or potentially using correlated-k \citep[e.g.][]{Amundsen2017}), at which point other effects may dominate as the principal sources of model uncertainty. Such effects include potential inaccuracies in molecular and atomic line lists, or indeed in the broadening of these lines as discussed above in the case of Na and K. The assumption of constant-with-depth chemical abundances may also impact model uncertainties, though in the case of T-dwarfs the abundances of H$_2$O and CH$_4$ (the dominant absorbers) are typically found to be constant with depth in equilibrium models \citep[e.g.][]{Line2017}. The presence of clouds and any horizontal structure (including patchy clouds) could also be important, especially for objects with known variability \citep[e.g.][]{Apai2013,Buenzli2013}. In future work, consideration of these effects could lead to an increasingly comprehensive understanding of model uncertainties and the potential to interpret brown dwarf spectra with ever increasing precision and accuracy.

\subsection{Generating Simulated Data}
\label{sec:sim_data}
As described in section \ref{sec:model}, we investigate different aspects of our model using a simulated data set based on Gl~570D. A simple way of doing this is to generate a reference model spectrum using the atmospheric model from the retrieval framework. However, such a spectrum would not factor in the model uncertainties described in section \ref{sec:model uncertainties} as, unlike real data, the simulated spectrum would be exactly described by the atmospheric model. In order to emulate real data as closely as possible and thereby test the retrieval framework's robustness to model uncertainties, we instead use the atmospheric model to generate the reference spectrum as accurately as possible, i.e. at the highest feasible resolution in both wavelength and pressure space. While this does not incorporate physical features that are not included in the atmospheric model (e.g. the presence of Na and K), the main numerical sources of uncertainty in the retrieval's atmospheric model are addressed and their effects on the retrieval can be tested.

We calculate the reference spectrum at the native resolution of the cross sections (0.1cm$^{-1}$), which is then binned in the same way as HST/WFC3 data. Since the real HST/WFC3 data that we analyse has uncertainties of 0.1-0.2\%, each binned model point is then shifted by a value randomly drawn from a Gaussian distribution with standard deviation equal to 0.2\% of the simulated data point's flux value. The simulated uncertainty is 0.2\% of the flux in the respective spectral bin. The radiative transfer calculation is done over 1000 layers (compared to 100 in the retrieval). Note that the simulated data is generated with the same pressure limits as the model in the retrieval framework and no Na or K is included, so uncertainties due to these factors are not present.

\subsection{Treatment of Uncertainties}

\label{sec:tol}

We have considered the magnitude of uncertainty in our model spectra in section \ref{sec:model uncertainties}. We now wish to include this information in the retrieval such that the retrieved atmospheric parameters are not biased by the model uncertainties. In addition, we would also like to allow for any unknown sources of error which may affect either the data or the model. This can be achieved by adding a `tolerance' parameter to the model which characterises the `extra' uncertainty in addition to the known observational uncertainties. Previous studies have done this by adding a wavelength-independent parameter in quadrature to the uncertainty of each data point \citep[e.g.][]{Line2015,Burningham2017}. The $\sigma_i$ term in equation \ref{eq:likelihood} that \citet{Line2015} use is given by:
\begin{equation}
\label{eq:ctol}
\sigma_i = \sqrt{\sigma_{i,d}^2 + 10^b},
\end{equation}
where $\sigma_{i,d}$ is the uncertainty in the $i^{\mathrm{th}}$ data point and $b$ is a free parameter with a uniform prior. The prior bounds on $b$ are such that the minimum possible value of $10^b$ is very small (typically less than the smallest error bar in the data), meaning that the retrieval is not obliged to account for the known model uncertainty. Otherwise put, this form of the tolerance parameter relies on the retrieval being able to ascertain the model uncertainty without it being included in the prior information. We test whether this can work by implementing this form of the tolerance parameter in our retrieval framework, with the uniform prior on $b$ going from log$_{10}(0.01 \, \sigma^2_{d, min})$ to log$_{10}(10000 \, \sigma^2_{d, max})$, where $\sigma^2_{d, min}$ and $\sigma^2_{d, max}$ are the minimum and maximum uncertainties in the data.

Figure \ref{fig:tolcomp} shows the posterior distributions of a selection of model parameters from a retrieval that uses this tolerance parameter (in purple). The retrieved value of $b$ is equivalent to adding an uncertainty of $10^b = 6.5\times 10^{-23}$ Wm$^{-2}$m$^{-1}$, i.e. many orders of magnitude smaller than the estimated $\sim$8\% uncertainty in the model ($10^b < 2 \times 10^{-10}$\% of the minimum flux value in the simulated data). This underestimate in uncertainty is manifested as very narrow posterior distributions which, for parameters such as the H$_2$O abundance, result in the median retrieved value being consistently offset from the true value by more than one standard deviation. Effectively, the `noise' introduced by the known model uncertainty is easily overfitted by a slight offset in the model parameter values. This can be remedied by explicitly accounting for known model uncertainties in the tolerance parameter, as discussed below.

Another potential issue with the tolerance parameter in equation  \ref{eq:ctol} is that it is wavelength-independent, whereas the model uncertainty is not. From figures \ref{fig:rescomp} and \ref{fig:NDcomp}, the uncertainties from cross section sampling and the number of atmospheric layers used are, to within a few percent, proportional to flux (for a sufficiently large number of atmospheric layers and cross section sampling resolution). We therefore investigate a new form of the tolerance parameter:
\begin{equation}
\label{eq:mtol}
\sigma_i = \sqrt{\sigma_{i,d}^2 + x_{tol}^2f_{i,m}^2},
\end{equation}
where $f_{i,m}$ is the value of the $i^{\mathrm{th}}$ model binned point and changes in each iteration of the nested sampling algorithm, and $x_{tol}$ is a free parameter with a uniform prior. In order to fully account for the model uncertainty that is known, the lower bound of this prior is chosen to be the estimated level of uncertainty of the model, i.e. $\sim$8\%. The upper bound of the prior on $x_{tol}$ is arbitrarily set to the large value of 100\%. Using this form for the tolerance parameter in the retrieval framework results in the posteriors shown in figure \ref{fig:tolcomp} (in blue). As expected, the posteriors are much wider and the atmospheric parameters are successfully retrieved within 1 standard deviation of the true value. We therefore choose to use the $x_{tol}$ tolerance parameter in the work that follows. Note that when retrieving real data, unknown uncertainties (taken into account by the $x_{tol}$ parameter) may be of the order of 8\% or more. In this case, imposing a lower limit of 8\% on $x_{tol}$ does not make much difference and the retrieved posteriors are similar for both of the aforementioned uncertainty consideration methods. Nevertheless, it is worth accounting for the model uncertainty in case the unknown uncertainties are smaller.

\section{A P-T model for Brown Dwarfs}

\label{sec:PT}
In this section we present a new parametric $P$-$T$ model for brown dwarfs, which comprises of multiple atmospheric layers and is parameterised according to the changes in temperature across those layers, as well as a fiducial temperature. While several parametric $P$-$T$ models are available for exoplanet atmospheres, we consider a new $P$-$T$ profile for the following reasons: firstly, isolated brown dwarfs have much steeper temperature profiles compared to irradiated planets, and changes in the $P$-$T$ gradient can have significant effects on the emergent spectrum. This, combined with the high precision of brown dwarf spectra, means that the $P$-$T$ model must be adequately flexible such that the parameter space of $P$-$T$ gradients can be properly explored in the retrieval. Secondly, since the thermal profiles of brown dwarfs are so different to those of irradiated planets, the photospheres of brown dwarfs can occur at different pressure ranges compared to planets. As such, it is important that the $P$-$T$ model is able to capture this and provide flexibility in the photosphere, which is the atmospheric region best constrained by the spectrum. We begin by exploring these properties of brown dwarfs in section \ref{sec:depthorigin}, with the purpose of informing our $P$-$T$ model.  In section \ref{sec:PTmodel}, we present the $P$-$T$ model and discuss the challenges of developing a $P$-$T$ parameterisation which satisfies the above requirements given known numerical artefacts which have been encountered in existing prescriptions.

\subsection{Thermal and Opacity Structure of Brown Dwarfs}
\label{sec:depthorigin}
\begin{figure*}
\centering
    	\includegraphics[width=0.8\textwidth]{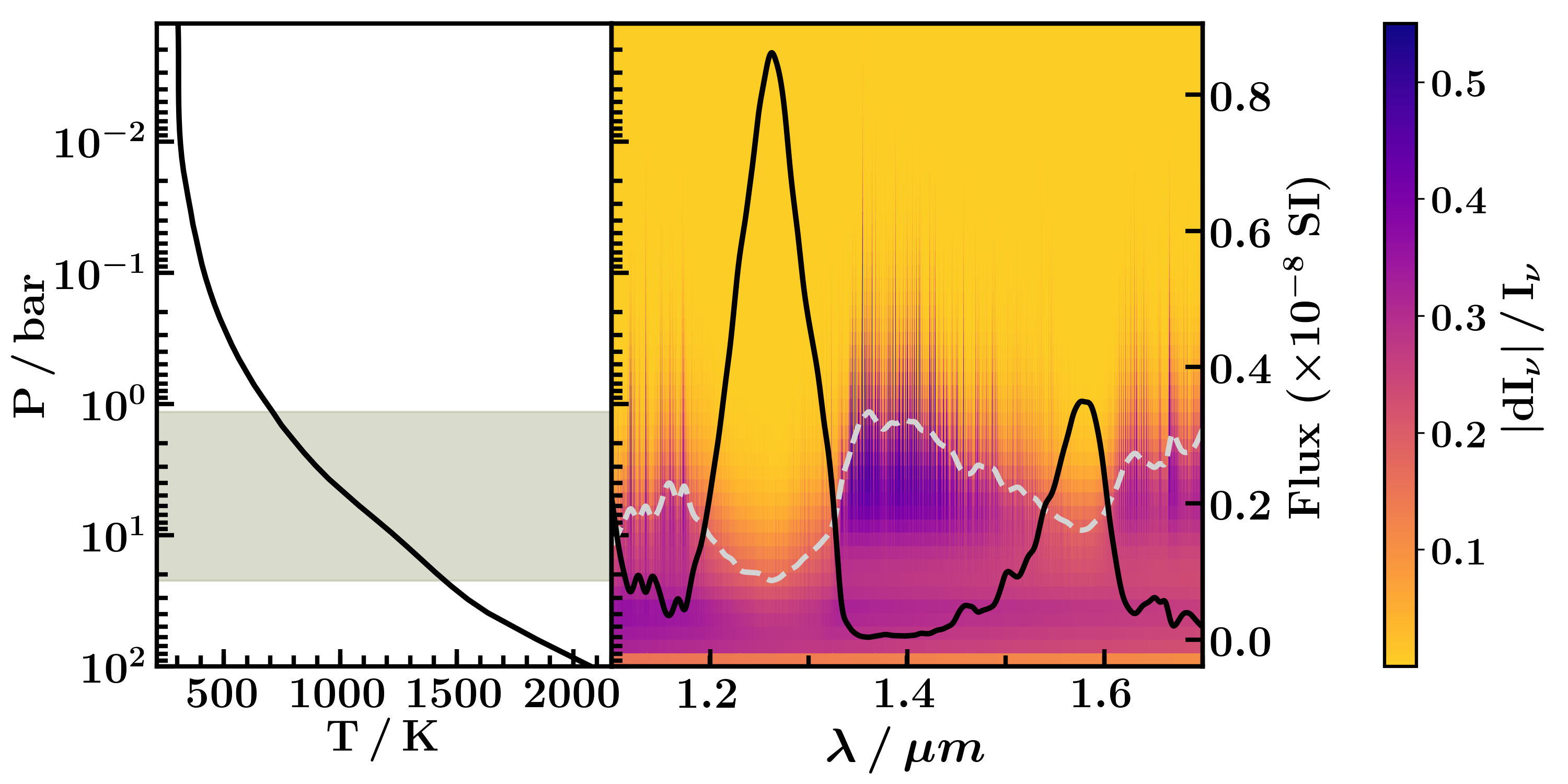}
\caption{\label{fig:sim_data_depth_origin} Left panel: approximation of the $P$-$T$ profile of G1 570D, based on the equilibrium profile from \citep{Line2015}. The range of pressures probed by the $\tau=1$ surface is shaded in gray. Right: colour map shows fractional change in spectral intensity, $|\textrm{dI}_{\nu}|/\textrm{I}_{\nu}$, where darker regions denote more absorption. The solid black line is the smoothed emergent spectrum for this atmosphere. The smoothed $\tau=1$ surface is shown by the dashed gray line and is indicative of the pressure to which a given part of the spectrum is most sensitive to. Flux axis is shown in SI units (Wm$^{-2}$m$^{-1}$).}
\end{figure*}

\begin{figure*}
\centering
    	\includegraphics[width=0.8\textwidth]{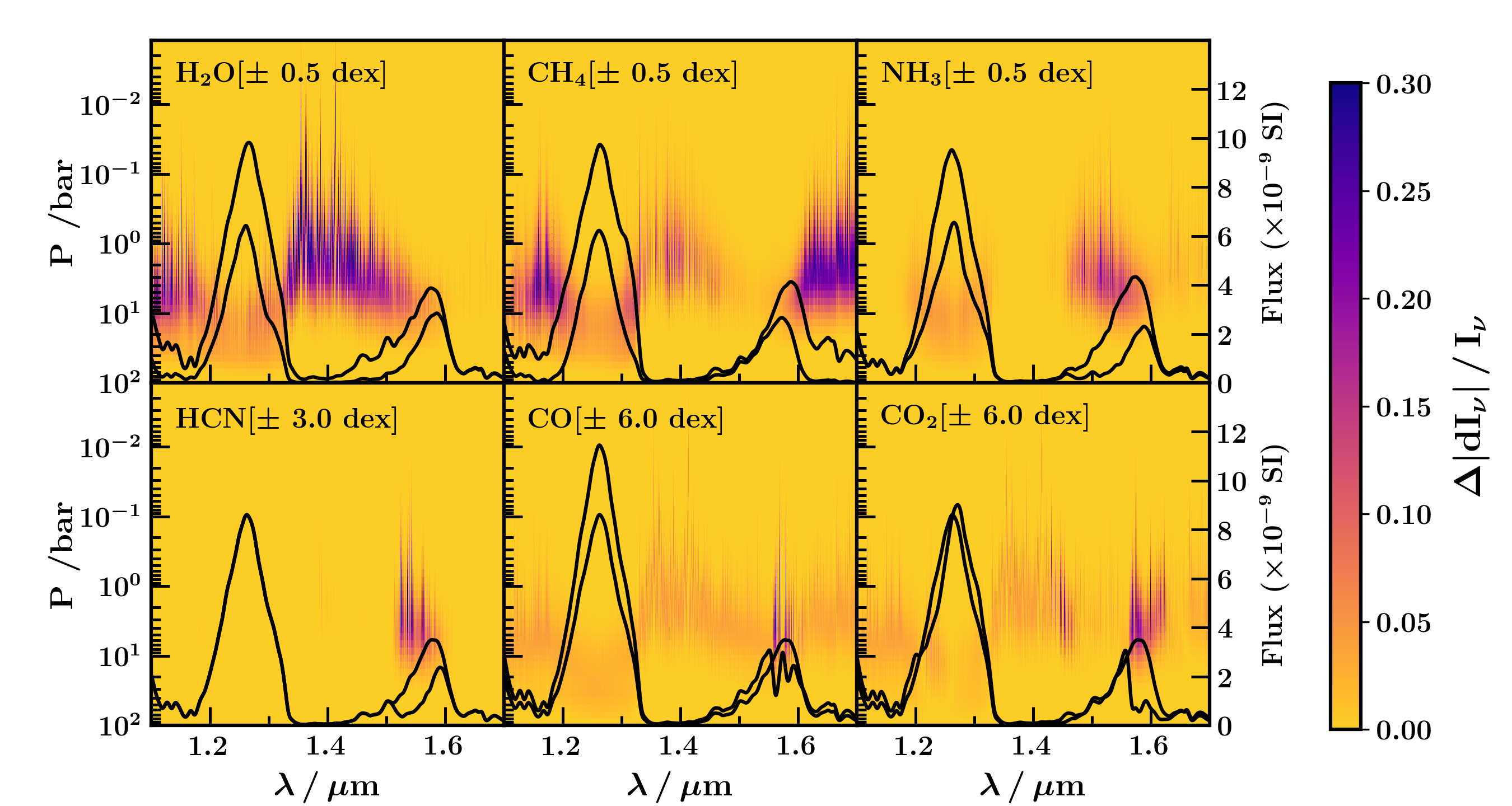}
\caption{\label{fig:sim_data_mol} Contributions of each species to atmospheric absorption. We take the synthetic spectrum of an atmosphere described by table \ref{tab:sim_data}, and perturb each species' abundance by the amount stated in the top left corner of each panel. The two black lines in each panel show the smoothed spectra with the positive and negative perturbations, respectively. Each colour map is the difference in $|\textrm{dI}_{\nu}|/\textrm{I}_{\nu}$ between the positive and negative perturbations. Note that larger perturbations in the abundances of HCN, CO and CO$_2$ (compared to H$_2$O, CH$_4$ and NH$_3$) are needed to make comparable changes in the spectrum. Note that HCN has been included in the reference model with an arbitrary mixing ratio of $10^{-7}$, and all other abundances are the same as in table \ref{tab:sim_data}. Flux axes are shown in SI units (Wm$^{-2}$m$^{-1}$).}
\end{figure*}

\begin{figure}
\centering
    	\includegraphics[width=0.5\textwidth]{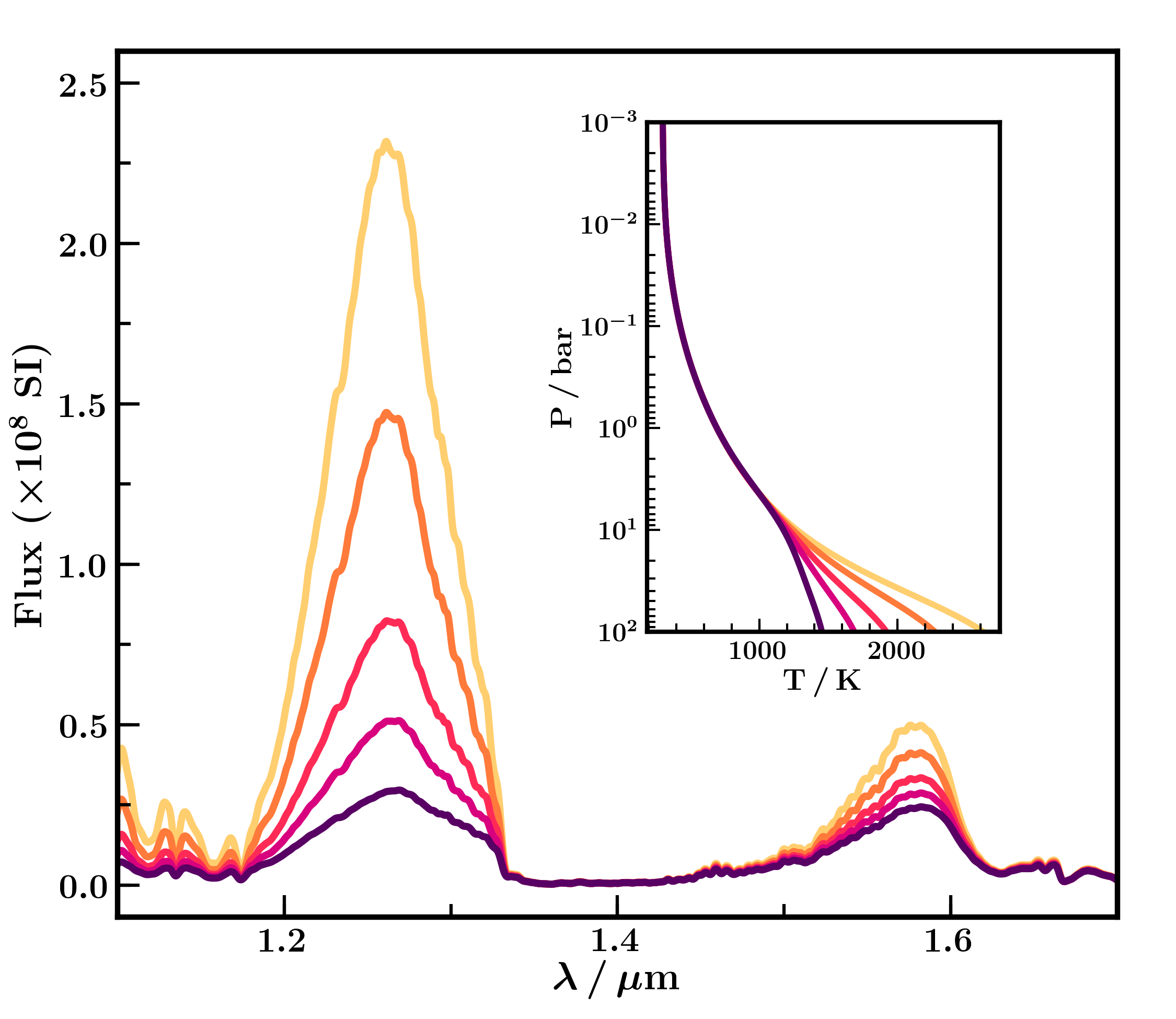}
\caption{\label{fig:PTslope} Effect of changing the slope of the $P$-$T$ profile at high pressures on the spectrum. Each $P$-$T$ profile in the inset corresponds to the spectrum with the matching colour. The slope of the $P$-$T$ profile at pressures greater than 10 bar has a significant effect on the flux peaks in the spectrum. Flux axis is shown in SI units (Wm$^{-2}$m$^{-1}$).}
\end{figure}

\begin{figure*}
\centering
    	\includegraphics[width=0.8\textwidth]{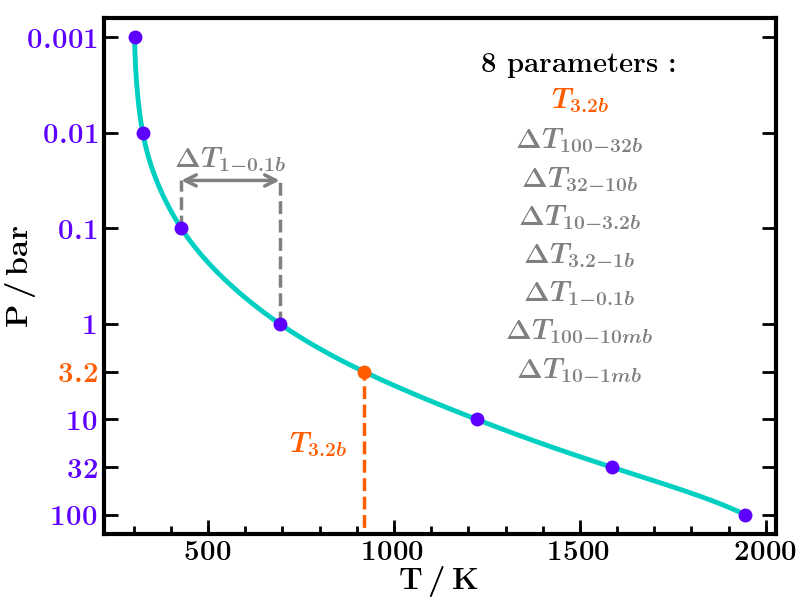}
\caption{\label{fig:PTcartoon} `SPT' parametric $P$-$T$ model used in our retrieval framework. The $P$-$T$ points are defined by the changes in temperature between them, as well as the absolute temperature at 3.2 bar. They are then interpolated using a monotonic spline fit and smoothed with a Gaussian kernel of width 0.3 dex in pressure.}
\end{figure*}
The majority of retrieval analyses to date have been performed on hot Jupiters. However, the thermal structure of an isolated brown dwarf differs greatly from that of a hot Jupiter as it does not receive stellar irradiation. Thus, these objects are not expected to exhibit thermal inversions or to have isotherms deep in their atmospheres ($\sim$ 1 bar), as found in hot Jupiters \citep{Hubeny2003, Fortney2008}. This has implications on how deeply we can probe the atmosphere of a brown dwarf; in known hot Jupiters, the isotherm at $\sim 0.1-1$ bar \citep{Gandhi2017} means that the spectrum is not sensitive to the atmosphere below this point. However, this limit does not exist for isolated objects, whose temperatures increase monotonically with pressure. Instead, the maximum depth which we are sensitive to in a brown dwarf is limited by the height at which the atmosphere becomes optically thick.

To investigate how a brown dwarf atmosphere is probed by its spectrum, we consider the origins of the spectrum as a function of pressure and wavelength. To quantify this, we define the fractional change in spectral intensity, $|\textrm{dI}_{\nu}|/\textrm{I}_{\nu}$, where $\textrm{I}_{\nu}$ is the incoming spectral intensity at the bottom of an atmospheric layer, and $\textrm{dI}_{\nu}$ is the change in the spectral intensity across that layer. We plot this quantity for our simulated model of G1 570D in figure \ref{fig:sim_data_depth_origin}, showing the depths which each point in the spectrum is sensitive to. The line of unit optical depth, $\tau=1$, represents the depth which the spectrum at a given wavelength is most sensitive to. In the case of this model atmosphere, we find that the spectrum is highly sensitive to pressures between 1 and 20 bar, and significantly affected by pressures up to several tens of bar.

The implications of such a deep photosphere are that (i) we are able to probe T-dwarf atmospheres much more deeply than hot Jupiters; (ii) the choice of parametrisation for the P-T profile should allow for a lot of flexibility between $\sim 1$ and $\sim 100$ bar, as the spectrum is able to strongly constrain this part of the profile. It is also informative to see how each species in the model contributes to the the absorption map in figure \ref{fig:sim_data_depth_origin}.
Figure \ref{fig:sim_data_mol} shows these contributions, from which it is clear that H$_2$O, CH$_4$ and NH$_3$ have the strongest features in this spectral range and are most likely to be constrained by the retrieval. It is also clear that there are degeneracies between the contributions of each molecule, although high-quality data is able to distinguish between their individual fingerprints.

Degeneracies also exist between the abundances of certain species and aspects of the $P$-$T$ profile. This can be understood by considering the effects of the slope of the $P$-$T$ profile on the spectrum, as shown in figure \ref{fig:PTslope}. The slope of the deepest part of the $P$-$T$ profile has a significant effect on the high-flux regions of the spectrum, which is expected since those parts of the spectrum are formed in the high-pressure regions of the atmosphere. These flux peaks are also strongly shaped by the abundances of the most active species in the atmosphere (in this case, H$_2$O, CH$_4$ and NH$_3$). As a result, the slope of the deepest part of the $P$-$T$ profile is degenerate with the abundances of H$_2$O, CH$_4$ and NH$_3$. Flexibility in the high-pressure region of the $P$-$T$ model is therefore crucial, as any deviation from the true $P$-$T$ profile may be propagated to the retrieved chemical abundances.

\subsection{P-T Parametrisation}

\label{sec:PTmodel}
In existing work on the retrieval of exoplanet atmospheres, a common approach has been to describe the $P$-$T$ profile with an analytic model \citep[e.g.][]{Madhusudhan2009, Guillot2010}. Studies using brown dwarf spectra \citep[e.g.][]{Burningham2017} have also considered such profiles \citep[e.g. that of][]{Madhusudhan2009}. These analytic models are well suited to atmospheric retrievals as they are able to capture a broad variety of $P$-$T$ profiles with only a few parameters. Indeed, they have been proven to work well with the current data quality of exoplanet spectra \citep[e.g.][]{Blecic2017,Gandhi2018}. However, this simplicity can come at a cost as an analytic function may not always be capable of exactly capturing the underlying $P$-$T$ profile, particularly for very high-quality data. In the case of T-dwarfs, an incorrect $P$-$T$ slope could potentially introduce biases into other retrieved parameters, as discussed in section \ref{sec:depthorigin}.

A second approach which has been used for retrievals of isolated or poorly-irradiated sub-stellar objects is to use physically-motivated non-irradiated $P$-$T$ profiles. For example, one of the models considered by \citet{Kitzmann2020} is an approximate solution to radiative transfer assuming a gray atmosphere and radiative equilibrium. They find that such a solution struggles to fit the temperature profile at lower pressures. \citet{GravityCollaboration2020} use a parametric $P$-$T$ model which applies the Eddington approximation in the photosphere and a moist adiabat below the radiative-convective boundary.

An alternative way of describing the $P$-$T$ profile is to allow the temperatures at certain pressures in the atmosphere to be free parameters, which can then be interpolated to the resolution at which radiative transfer is calculated \citep[e.g][]{Line2015,Line2017,Zalesky2019,Kitzmann2020}. This form of parametrisation has the advantage of allowing a lot of freedom in the $P$-$T$ profile, but this freedom is susceptible to over-fitting. In particular, oscillations in the $P$-$T$ profile (or `ringing') are a common problem. \citet{Line2015,Line2017} overcome this by penalising the second derivative of the $P$-$T$ profile in the likelihood function, disfavoring features in the profile which are not required by the data. \citet{Kitzmann2020} consider a different approach and fit the $P$-$T$ profile with a piecewise polynomial.

We present a new approach to describing the $P$-$T$ profile which allows for a high level of freedom but excludes unphysical features. In particular, we aim to avoid ringing and large oscillations which resemble thermal inversions, as these can easily overfit the data but are evidently not physical. Large oscillations in the $P$-$T$ profile rely on the freedom to have temperature increasing towards lower pressures, and so one way to avoid this is to exclude $P$-$T$ models with a negative gradient at any point. This can be done by assigning a very low likelihood to any models with a temperature profile which does not monotonically increase with pressure. However, this typically results in a very low acceptance rate in the nested sampling algorithm and hinders its convergence. To overcome this, we instead characterise the temperature profile by the changes in temperature, $\Delta T_i$,  across multiple atmospheric layers, plus a temperature parameter which acts as an anchor point (`slope' P-T model, or `SPT'; see figure \ref{fig:PTcartoon}). The temperature parameter is placed in the photosphere as this is where the $P$-$T$ profile is best constrained by the data. Based on figure \ref{fig:PTslope}, the range of pressures covered by the photosphere is expected to include $\sim$1-10 bar, so we place the temperature parameter at 3.2 bar ($T_{\mathrm{3.2 b}}$).

In each atmospheric layer, $\Delta T_i$  is the increase in temperature from the low-pressure side of the layer to the high-pressure side. We therefore set the lower bound of the prior distributions on the $\Delta T_i$ to 0K such that only models where temperature increases with pressure are considered. The upper bounds of the priors are shown in table \ref{tab:priors}. The prior for each $\Delta T_i$ parameter is chosen to be a very generous temperature range given typical $P$-$T$ profiles of T-dwarfs. We avoid placing unrealistically high limits on the $\Delta T_i$ priors as this would result in many models extending to unfeasibly high or low temperatures. Instead, we choose a suitably wide prior for each individual $\Delta T_i$ parameter.

For given $\Delta T_i$ and $T_{\mathrm{3.2 b}}$, the temperature nodes defined by these parameters are interpolated with a monotonic spline and smoothed with a Gaussian kernel of width 0.3 dex in pressure. Note that we do not use a `traditional' spline fit as this is prone to introducing oscillations, even if the nodes it is fitting are monotonically decreasing. The number of $\Delta T_i$ parameters and the thicknesses of the atmospheric layers between them are free to be chosen -  we use the 7 layers defined between pressures of 100, 32, 10, 3.2, 1, 0.1, 0.01 and 0.001 bar as this concentrates the $\Delta T_i$ parameters at deeper pressures, where more flexibility is needed, and has fewer parameters at lower pressures, where the spectrum does not strongly constrain the $P$-$T$ profile. We note, however, that this $P$-$T$ model is not strongly sensitive to the chosen atmospheric layers as long as enough flexibility is given in the photosphere.

\section{Retrieval of Simulated Data}

\begin{figure*}
\centering
    \begin{subfigure}[b]{0.48\textwidth}
  		\includegraphics[width=\textwidth]{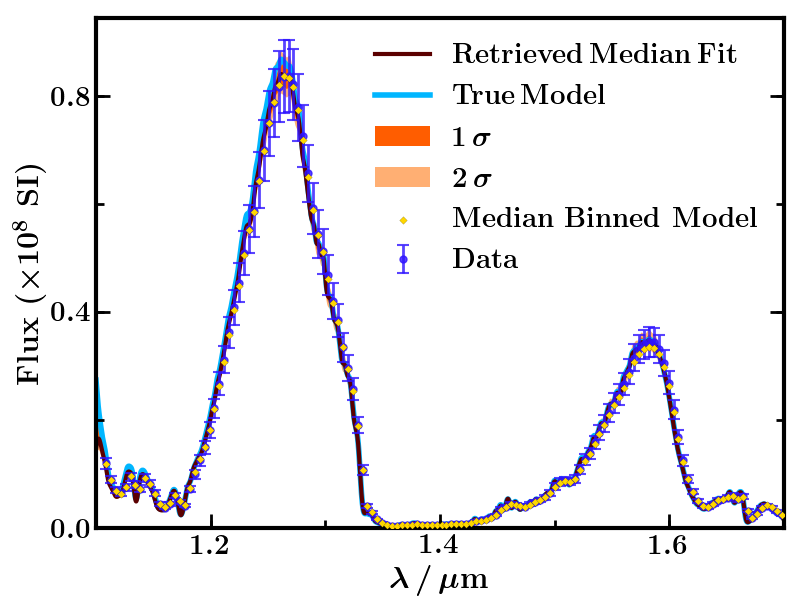}
    \end{subfigure}
    \begin{subfigure}[b]{0.48\textwidth}
    	\includegraphics[width=\textwidth]{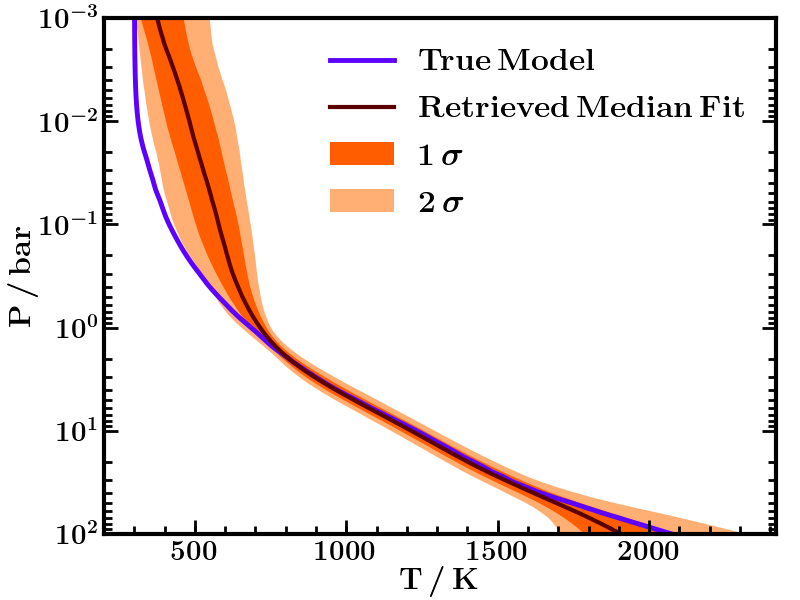}
    \end{subfigure}
\caption{\label{fig:sim_spec} Left: The median retrieved spectrum (maroon line) from the retrieval of our simulated data. 1$\sigma$ and 2$\sigma$ confidence intervals are shown by dark and light orange shading, respectively. The simulated data points are shown by blue circles, and error bars show the median retrieved tolerance. Yellow diamonds show the median binned model. Flux axis is shown in SI units (Wm$^{-2}$m$^{-1}$). Right: The input (blue line) and median retrieved (maroon line) $P$-$T$ profiles. 1$\sigma$ and 2$\sigma$ confidence intervals are shown by dark and light orange shading, respectively.}
\end{figure*}

\begin{figure*}
\centering
    	\includegraphics[width=0.9\textwidth]{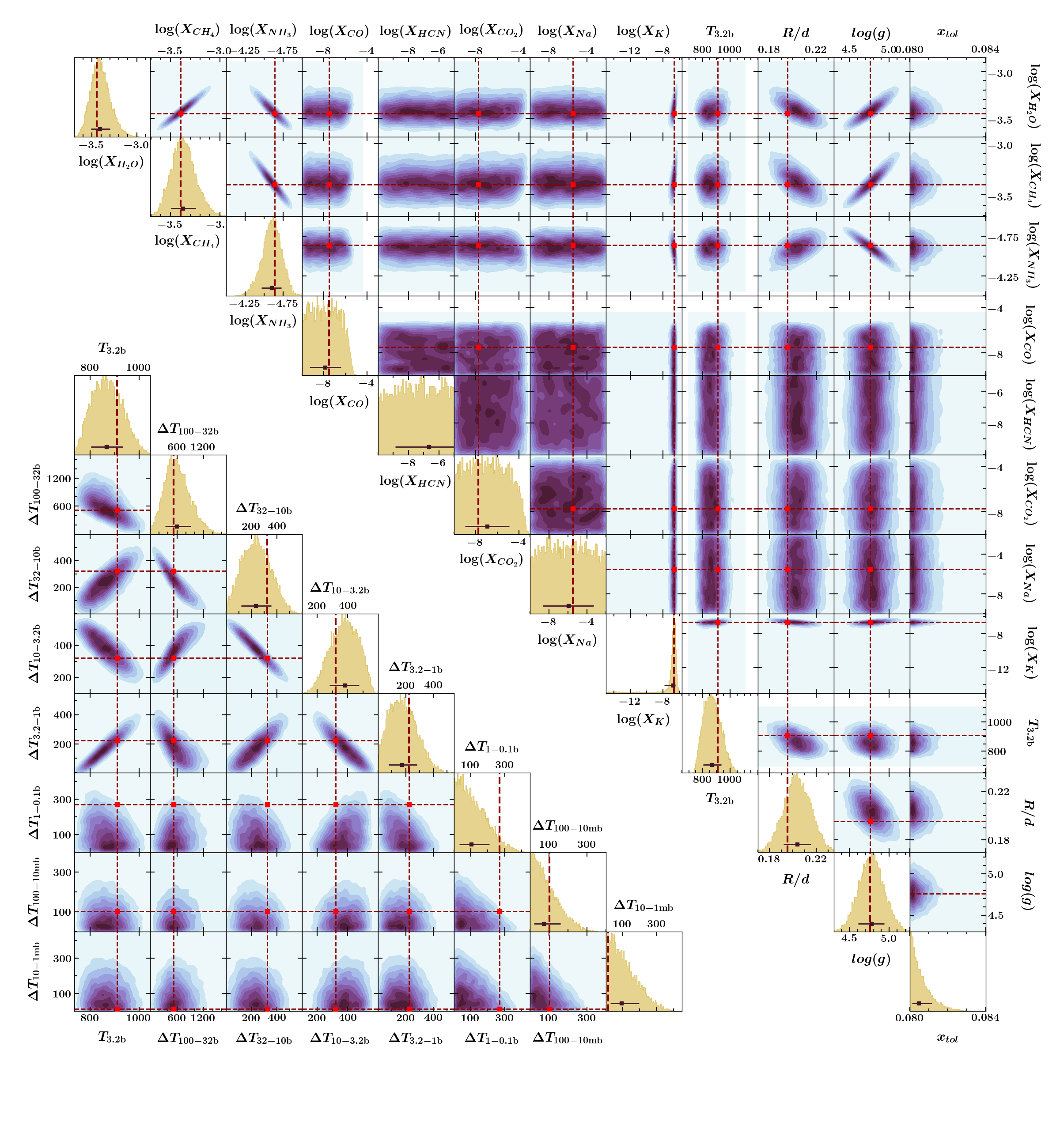}
\caption{\label{fig:sim_corner} Marginalised posterior probability distributions from the retrieval of simulated data based on Gl~570D. Off-axis plots are 2D marginalised posteriors and show the correlations between pairs of parameters. 1D marginalised posteriors are shown for each parameter along the diagonals, with median values and 68\% confidence intervals shown by the dark squares and error bars, respectively. Dashed lines show the input values of each parameter, used to generate the simulated data.}
\end{figure*}

\label{sec:self-consistency}
We now present a test retrieval on our simulated data (described in section \ref{sec:sim_data})  using the chosen tolerance parameter and the SPT $P$-$T$ model from sections \ref{sec:uncertainties} and \ref{sec:PT}, respectively. There are a total of 19 parameters: 8 chemical abundances, 8 $P$-$T$ parameters, gravity, radius-to-distance ratio and the tolerance parameter. The cross sections we use for Na and K include the wings of the strong optical lines (at 0.6 $\mu$m and 0.8 $\mu$m, respectively) as well as the other Na and K lines present in the spectral range 1.1-1.7 $\mu$m \citep{NIST_ASD}. We use the broadening profiles of \citet{Allard2016} and \citet{Allard2019} for all of the lines, scaled appropriately according to line strength. We run this retrieval on the simulated data described in section \ref{sec:sim_data}, for which the input parameters are known.

The retrieved parameter values match the known parameter values excellently (figures \ref{fig:sim_spec} and  \ref{fig:sim_corner}, table \ref{tab:sim_data}). All of the constrained parameters are retrieved within 1$\sigma$ of the true values, and the simulated data points are successfully fitted by the median retrieved spectrum. In particular, the abundances of H$_2$O, CH$_4$, NH$_3$ and K as well as gravity and the radius-to-distance ratio are strongly constrained. The abundances of CO, HCN, CO$_2$ and Na are not constrained but this is expected since they do not have strong features in this spectral range. The retrieved $P$-$T$ profile in figure \ref{fig:sim_spec} is an excellent fit to the input profile in the photosphere, where it is most tightly constrained by the spectrum, though below pressures of $\sim$ 1 bar and above $\sim$ 30 bar the $P$-$T$ profile has little influence on the spectrum and is not tightly constrained. We therefore conclude that this retrieval framework is consistent, in the sense that it is able to accurately retrieve a high-precision brown dwarf spectrum generated using its parametric model.  Furthermore, the retrieval is able to successfully use a lower-resolution atmospheric model to retrieve a model spectrum generated with higher spectral and vertical resolution.

The test retrieval also highlights degeneracies which exist between certain pairs of parameters, although they do not prevent the retrieval from finding the true parameter values. It is informative to understand the origins of these degeneracies, which we describe below.   The strongest degeneracies between the parameters, visible in figure \ref{fig:sim_corner}, are between the chemical abundances, gravity and $R/d$.

\begin{enumerate}
\item \textbf{Gravity and Chemical Abundances: }Gravity and the chemical abundances are positively correlated because they both affect the optical depth of the atmosphere. Since pressure is an independent variable in our atmospheric model, the equation of hydrostatic equilibrium (equation \ref{eq:hydrostatic_eq}) scales the physical depth of each atmospheric layer across which radiative transfer is calculated.  A larger log($g$) results in a smaller scale height and a lower optical depth. Conversely, greater chemical abundances result in a greater optical depth (equation \ref{eq:dtau}). This can be expressed by combining equations \ref{eq:dtau} and \ref{eq:hydrostatic_eq}:
\begin{equation*}
\frac{\mathrm{d}P}{\textrm{d}\tau} = \frac{\rho g}{\sum_i \sigma_i n_i},
\end{equation*}
where the sum is over all species. The optical depth can be kept constant for an increasing $g$ if the following expression also increases:
\begin{equation*}
\frac{\sum_i \sigma_i n_i}{\rho} = \sum_i \frac{\sigma_i X_i}{\bar{m}}
\end{equation*}
where $\bar{m}$ is the average particle mass.  In a hypothetical spectral range in which only one species is absorbing, this can be achieved by increasing that species' abundance. However, for a spectral band in which multiple species are absorbing, those with stronger cross sections will have increased abundances while those with weaker cross sections will have decreased abundances (since $\sum_{i}X_{i}=1$). In this case, H$_2$O, CH$_4$ and K are the strongest absorbers in their respective bands so their abundances are positively correlated with log($g$). However, NH$_3$ has a much weaker effect on the spectrum and is negatively correlated with log($g$). Since different species dominate in different spectral bands, the effect of varying the chemical abundances has a different wavelength dependence to varying the gravity. As such, this degeneracy does not prevent the true abundances and gravity from being determined in the retrieval.\\

\item  $\mathbf{R/d}$ \textbf{and Gravity: }Gravity and $R/d$ are negatively correlated as they both scale the flux level of the spectrum. As discussed above, a larger value of log($g$) results in  a lower optical depth and a higher flux. Since $R^2/d^2$ is a direct multiplicative factor in the flux (equation \ref{eq:Fnu}), increasing it also results in a larger flux. An increase in log(g) can therefore be somewhat offset by a decrease in $R/d$, resulting in a negative correlation. Nevertheless, the wavelength dependences of these effects are different, meaning that this is not a strong degeneracy and is broken in the retrieval. \\

\item  $\mathbf{P-T}$\textbf{ parameters: }The $\Delta T$ parameters deeper than 1bar are all degenerate with each other and show an interesting pattern of negative correlations between adjacent atmospheric layers and positive correlations between $\Delta T$s separated by an odd number of layers. This pattern is indicative of oscillations in the $P$-$T$ profile and is significantly accentuated by the smoothing of the profile, as profiles with small oscillations become equivalent to ones without them.

\item  \textbf{Chemical Abundances: }The abundances of some of the constrained species (H$_2$O, CH$_4$ and K) are strongly positively correlated. If the spectral fingerprints of these molecules were not entirely distinguishable, one would expect a negative correlation between their abundances. However, the data is able to distinguish each species and the degeneracy instead comes from the fact that the chemical abundances are degenerate with bulk properties of the atmosphere such as gravity and $R/d$. For example, an increase in the abundance of H$_2$O can be offset by changes in log($g$) or $R/d$, but these quantities are also degenerate with the abundances of CH$_4$ and K which must then also increase.
\end{enumerate}

\section{Application to Data}
\label{sec:results}
\begin{table}
    \centering
    \begin{tabular}{c|c|c}
         Property & Value & References\\
         \hline
         RA (deg, J2000)& 354.792715 & 1,2\\
         DEC (deg, J2000)& +13.874577 & 1,2\\
         IR Spectral Type & T5 & 3\\
         Estimated distance (pc) & 18.8$\pm$3.8 & 4\\
         2MASS J mag & 16.239$\pm$0.108 & 1,2\\
         2MASS H mag & 15.822$\pm$0.151 & 1,2\\
         W1 mag & 15.218$\pm$0.044 & 5\\
         W2 mag & 13.818$\pm$0.044 & 5\\
         $\mu_\alpha$ (mas yr$^{-1}$) &  396.0$\pm$41.6 & 6 \\
         $\mu_\delta$ (mas yr$^{-1}$) &  -1018.7$\pm$34.6  & 6 \\

    \end{tabular}
    \caption{Properties of 2MASS J23391025+1352284 from the literature. \textbf{References:} 1. \citet{Cutri2003}, 2. \citet{Skrutskie2006}, 3. \citet{Burgasser2006}, 4. \citet{Buenzli2014}, 5. \citet{Wright2010}, 6. \citet{Schneider2016} }
    \label{tab:BDprops}
\end{table}

\begin{figure*}
\centering
    \begin{subfigure}[b]{0.48\textwidth}
  		\includegraphics[width=\textwidth]{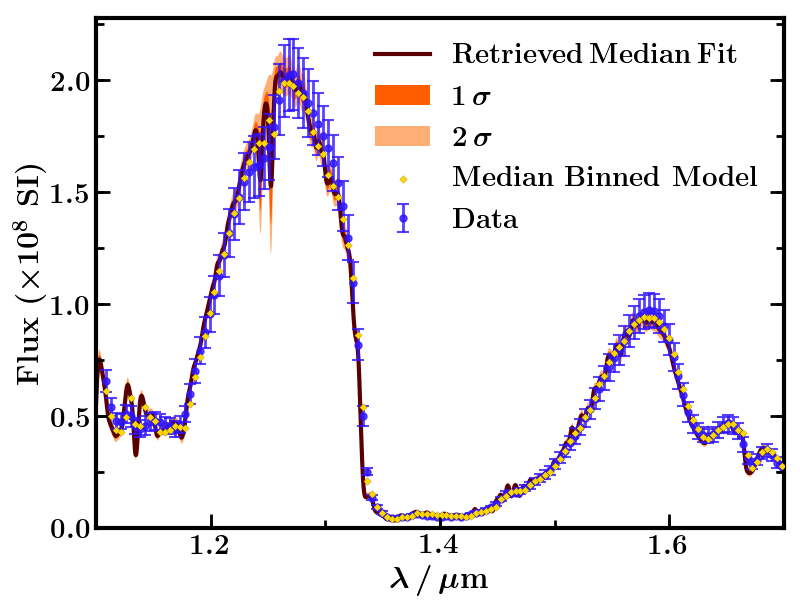}
    \end{subfigure}
    \begin{subfigure}[b]{0.48\textwidth}
    	\includegraphics[width=\textwidth]{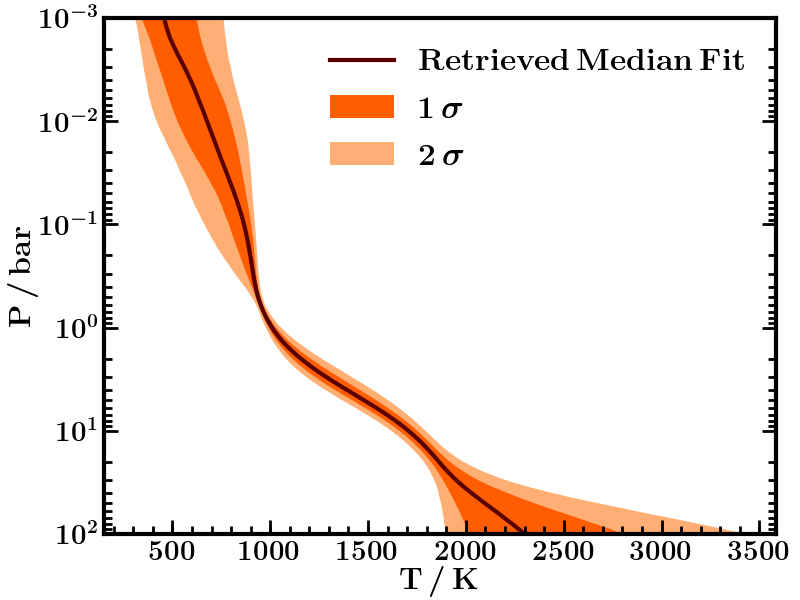}
    \end{subfigure}
\caption{\label{fig:data_spec} Spectrum and $P$-$T$ profile retrieved for 2MASS J2339+1352. Left: The median retrieved spectrum (maroon line) and 1$\sigma$ and 2$\sigma$ confidence intervals (dark and light orange shading, respectively). Data points are shown by blue circles, and error bars depict the median retrieved tolerance. Yellow diamonds show the median binned model. Flux axis is shown in SI units (Wm$^{-2}$m$^{-1}$). Right: Median retrieved $P$-$T$ profile (maroon line) and 1$\sigma$ and 2$\sigma$ confidence intervals (dark and light orange shading, respectively).}
\end{figure*}
\begin{figure*}
\centering
    	\includegraphics[width=0.9\textwidth]{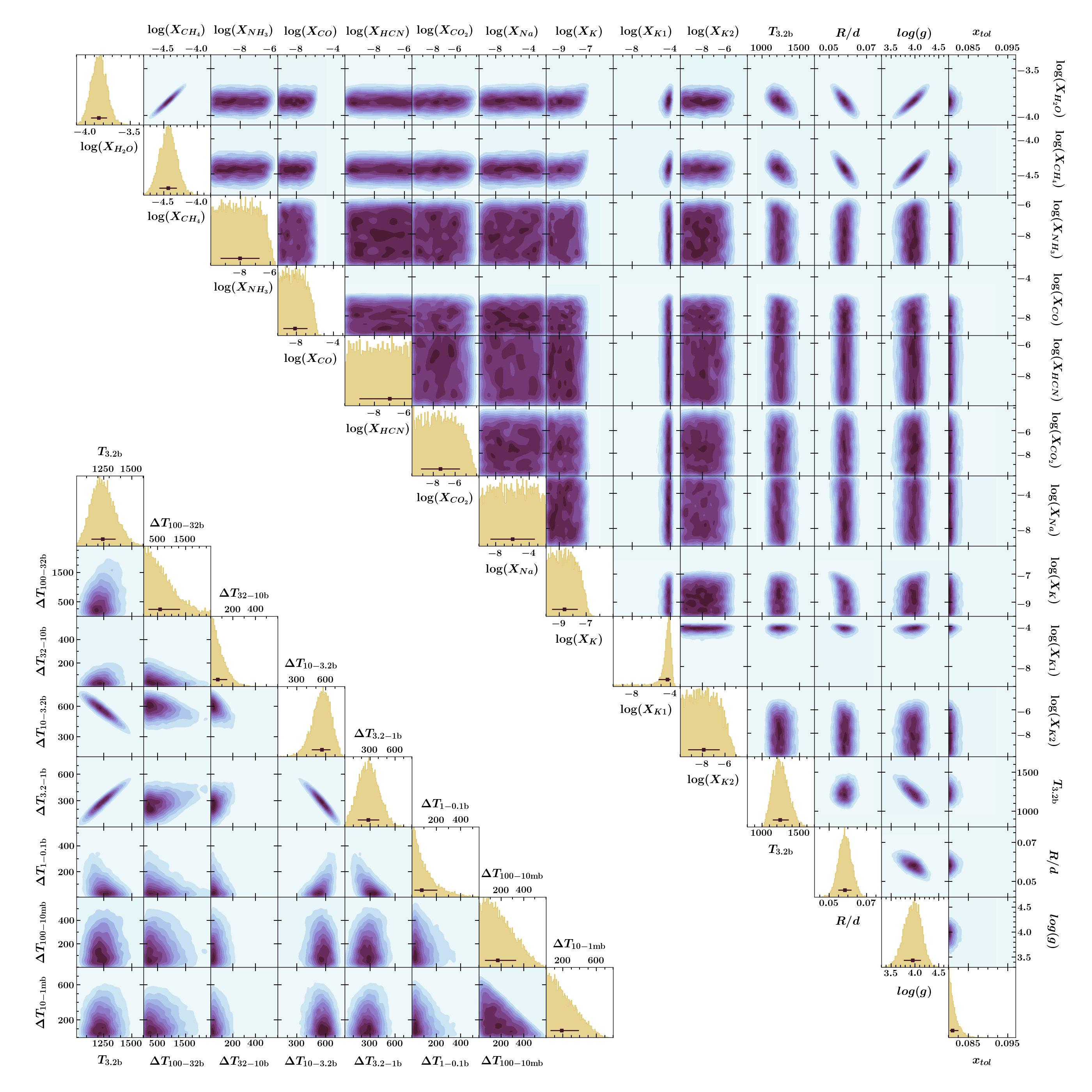}
\caption{\label{fig:data_corner} Marginalised posterior probability distribution from the retrieval of 2MASS J2339+1352. Off-axis plots show 2D marginalised posteriors. 1D marginalised posteriors are shown for each parameter along the diagonals, with median values and 68\% confidence intervals shown by the dark squares and error bars, respectively.}
\end{figure*}

\begin{figure*}
\centering
    	\includegraphics[width=0.7\textwidth]{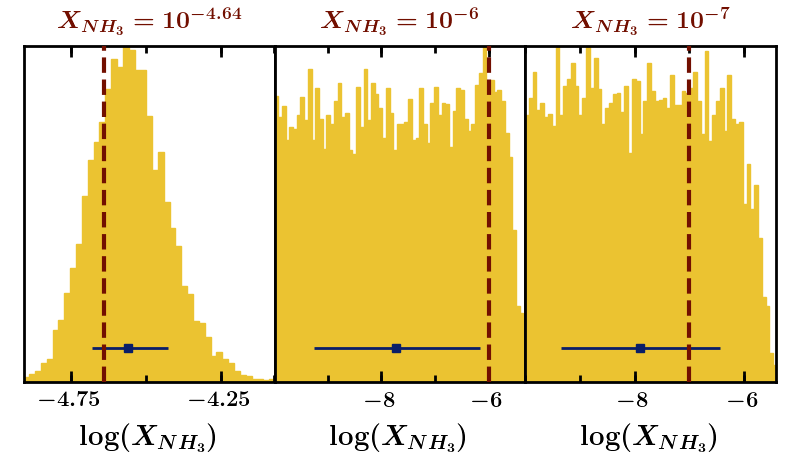}
\caption{\label{fig:nh3comp} Posterior distributions of the NH$_3$ abundance from the retrieval of simulated data with input NH$_3$ mixing fractions of $10^{-4.64}$, $10^{-6}$ and $10^{-7}$, respectively. Median retrieved values and 68\% confidence intervals shown by the dark blue squares and error bars, respectively. Vertical dashed lines show the true NH$_3$ abundance for each case. NH$_3$ is not detected when the mixing fraction is $10^{-6}$ or $10^{-7}$.}
\end{figure*}

We now demonstrate our retrieval framework on an emission spectrum of the brown dwarf 2MASS J23391025+1352284 \citep{Buenzli2014}. Properties of 2MASS J2339+1352 from the literature are shown in table \ref{tab:BDprops}. This object was discovered by the Two Micron All-Sky Survey (2MASS; \citealt{Cutri2003,Skrutskie2006}) and identified as a T-dwarf by \citet{Burgasser2002}. Its infrared spectral sub-type is T5 \citep{Burgasser2006} and it is therefore expected to have an effective temperature in the range $\sim$1000-1500~K \citep[e.g.][]{Kirkpatrick2005}. A high proper motion has also been measured for this object (\citealt{Schneider2016}; see table \ref{tab:BDprops}), consistent with its nearby estimated distance of 18.8$\pm$3.8~pc \citep{Dupuy2012,Buenzli2014}. \citet{Buenzli2014} observed 2MASS~J2339+1352 using the HST's Wide Field Camera 3 (WFC3) in the range 1.1-1.7 $\mu$m, which is the spectrum we use for this retrieval. These data are part of a spectroscopic survey searching for brown dwarf variability, which in turn may indicate the presence of patchy photospheric clouds. \citet{Buenzli2014} find that, in contrast to many of the brown dwarfs in their survey, 2MASS~J2339+1352 is only tentatively variable. This object is therefore a suitable demonstration for our cloud-free retrieval. The uncertainties on the data are of order 0.1\%, with an estimated 2\% systematic flux calibration uncertainty. The model uncertainties considered in section \ref{sec:uncertainties} are therefore significant and justify the use of the tolerance parameter.  As described in section \ref{sec:uncertainties}, we also parameterise the cross sections of the strongest K lines in this spectral range.

Figures \ref{fig:data_spec} and \ref{fig:data_corner} show the results of the retrieval. In what follows, we describe the retrieved values of the model parameters.

\subsection{Chemical abundances}

\textbf{H$_2$O and CH$_4$: }
We obtain strong detections of H$_2$O and CH$_4$ with log abundances of $-3.88 \pm 0.09$ and $-4.5 \pm 0.1$, respectively. The $\sim 0.1$ dex precisions of these abundances are smaller than the $\sim 0.5$ dex achievable with current exoplanet spectra \citep[e.g.][]{Gandhi2018} thanks to the precision of the data. Furthermore, these retrieved abundances are lower than expectations based on solar elemental abundances in chemical equilibrium \citep{Burrows1999,Madhusudhan2012}. Using H$_2$O and CH$_4$ as proxies for the abundances of oxygen and carbon, respectively, their abundances suggest that oxygen is $0.13\pm0.03$ times sub-solar and that carbon is $0.07^{+0.03}_{-0.02}$ times sub-solar \citep{Asplund2009}. However, we note that other sources of oxygen and carbon (e.g. silicates or CO) which are not detectable in the near-infrared spectrum would alter these estimates.

\textbf{NH$_3$: }
Although we found that NH$_3$ could be retrieved from our simulated data, we only retrieve an upper limit from the spectrum of 2MASS J2339+1352. This could potentially be due to a low abundance of NH$_3$, and to test this we create and retrieve two further sets of simulated data identical to that described in section \ref{sec:sim_data}, but with reduced NH$_3$ mixing ratios of $10^{-6}$ and $10^{-7}$, respectively. In both cases, we retrieve a flat posterior for the NH$_3$ abundance, indicating that NH$_3$ is not detected (figure \ref{fig:nh3comp}). It is therefore plausible that we are seeing signs of a very low abundance of NH$_3$ in the spectrum of 2MASS J2339+1352. This is consistent with the retrieved $P$-$T$ profile since NH$_3$ is not expected to be abundant at higher temperatures, when it is preferentially locked up in N$_2$; for example, the equilibrium solar mixing ratio of NH$_3$ at 1bar drops from $\sim 10^{-3.7}$ at 500K to $\sim 10^{-5.2}$ at 1000K \citep{Burrows1999, Moses2013}.

\textbf{CO, CO$_2$ and HCN: }
We retrieve an upper limit on the abundance of CO of $10^{-5.9}$ (99\% confidence interval). However, the abundances of CO$_2$ and HCN are unconstrained by the retrieval. This is expected since these species do not have strong cross sections in the spectral range of the data (figure \ref{fig:sim_data_mol}).

\textbf{Na and K: }
We place an upper limit on the K mixing ratio of $10^{-6.9}$ (at the 99\% confidence level). We also find that the spectral feature visible in the data at $\sim$1.25 $\mu$m is fitted by the K lines in this region, resulting in a strong constraint on $X_\mathrm{K1}$. $X_\mathrm{K2}$ is not constrained by the retrieval, suggesting that these lines are in fact too weak to be recovered from the spectrum. Na is completely unconstrained by the retrieval, as expected from the test retrieval in section \ref{sec:self-consistency}, as it has very little impact on the spectrum.

In order to test the effects of Na and K on the other retrieved parameters, we also performed a retrieval without Na and K (figures \ref{fig:nonak_spec} and \ref{fig:nonak_corner}, table \ref{tab:nonak}). The retrieved values of all the other parameters are consistent between the two retrievals, though the feature at $\sim$1.25 $\mu$m is not fitted when K is not included. This indicates that, while including the K lines at $\sim$1.25 $\mu$m improves the fit to the data, it does not impact the other retrieved parameters.

\subsection{P-T profile}
The retrieved $P$-$T$ profile is well constrained, especially in the photosphere. Between $\sim$0.6 and $\sim$20 bar, the temperature is constrained to within $\sim$100 K. We retrieve the temperature at 3.2 bar (i.e. the fiducial temperature in the photosphere, see section \ref{sec:PTmodel}) to be 1240$^{+110}_{-100}$ K. As expected, the temperature in the deepest and shallowest regions of the atmosphere are not well constrained by the retrieval, as can be seen by the larger 1$\sigma$ and 2$\sigma$ margins in the right panel of figure \ref{fig:data_spec}. This shows that the retrieval is able to fit parts of the $P$-$T$ profile which inform the spectrum while remaining agnostic about parts which are not constrained by the data.

\subsection{R/d, gravity and tolerance}
We retrieve a radius-distance ratio of 0.058$\pm$0.004 $R_\mathrm{J}$/pc and a log gravity (in cms$^{-2}$) of 4.0$\pm$0.2. Using a distance estimate, these quantities can be used to infer a mass and radius for 2MASS J2339+1352. Although a parallax measurement is not available for this object, \citet{Buenzli2014} calculate a distance estimate based on the relation between spectral type and absolute H-band magnitude given by \citet{Dupuy2012}. Using this estimate of 18.8$\pm$3.8 pc, we derive a radius of 1.1$\pm$0.2 $R_\mathrm{J}$ and a mass of 5$^{+3}_{-2}$ $M_\mathrm{J}$. We note, however, that a direct parallax measurement could provide more accurate mass and radius estimates. These mass and radius values are discussed further in section \ref{sec:discussion}.

The posterior probability distribution retrieved for the tolerance parameter is stacked against the lower prior of 8\%, suggesting that no significant uncertainties above this level were found by the retrieval. However, forcing the lower prior of this parameter to be the minimum known uncertainty in the model (i.e. $\sim$8\%) means that this uncertainty has been propagated to the posterior distributions of the other retrieved parameters, preventing underestimates in their uncertainties.

\begin{table}
\begin{centering}
\begin{tabular}{c|c|c}
Parameter & Retrieved value \\[3pt]
 \hline
 $\textrm{log}(X_{H_2O})$  & $-3.85 \pm 0.09$  \\[7pt]
 $\textrm{log}(X_{CH_4})$  & $-4.44 \pm 0.1$  \\[7pt]
 $\textrm{log}(X_{NH_3})$  & $<-5.9$  \\[7pt]
 $\textrm{log}(X_{\mathrm{CO}})$  & $ <-5.9 $  \\[7pt]
 $\textrm{log}(X_{\mathrm{K}})$  & $ <-6.9 $  \\[7pt]
 $T_{3.2\textrm{b}}$ (K) & $1240^{+110}_{-100}$  \\[7pt]
 $R/d\,\,(R_\mathrm{J}/\mathrm{pc})$  & $0.058 \pm 0.004$  \\[7pt]
 $\textrm{log}(g/\mathrm{cms}^{-2})$  & $4.0 \pm 0.2$  \\[7pt]
 $x_{tol}$ & $ 8.1^{+0.2}_{-0.07} \%$  \\[7pt]
\end{tabular}
\caption{\label{tab:real_data} A summary of the retrieved atmospheric properties of 2MASS J2339+1352. Upper limits show 99\% confidence intervals. The abundances of CO, HCN, CO$_2$, and Na are not constrained. The abundances of H$_2$O and CH$_4$ are both found to be significantly sub-solar, at $\sim$0.2$\times$ and $\sim$0.1$\times$ solar, respectively \citep{Gandhi2017}.}
\end{centering}
\end{table}

\section{Summary and Discussion}
\label{sec:discussion}
In this work we investigate important considerations for accurate atmospheric retrievals of high-precision spectra of brown dwarfs. This is motivated by the availability of very high-SNR HST infrared spectra of brown dwarfs. These data present the potential to determine the atmospheric and chemical properties of brown dwarfs with unprecedented precision. To utilise this potential we find that certain approaches in modelling technique contribute significantly to the accuracy of the results obtained. We introduce several key developments as follows.

Firstly, we consider the uncertainty in our atmospheric model and explicitly include it in the retrieval framework as prior information on a variable `tolerance' parameter. This allows the retrieval to constrain unknown sources of uncertainty \citep[e.g.][]{Line2017,Burningham2017}, while also accounting for known sources of error in the model. Two main sources of model uncertainty are the sampling of chemical cross sections and the resolution of the pressure grid in the model atmosphere. We approximate these effects as being proportional to model flux, and as such choose a tolerance parameter ($x_{tol}$) proportional to flux, which is then added in quadrature to the data uncertainties. The lower bound of the prior on $x_{tol}$ is then set to the level of known model uncertainties ($\sim$8\% in the present case). This error, which comes largely from our chosen spectral resolution, can be reduced by using a higher spectral resolution at a higher computational cost.  Although we find that using a tolerance parameter proportional to flux works well in estimating model uncertainties, other wavelength dependences are not accounted for and could be considered in future work. There are also potential sources of uncertainty in the model, such as depth-varying chemical abundances and atmospheric dynamics, which we do not include here but which could be considered in future work.

The second development we make is a new parametrisation of the $P$-$T$ profile. We begin by investigating the thermal and opacity structure of brown dwarfs in order to inform our choice of $P$-$T$ model and find that pressures of up to several tens of bar can be probed by the spectrum; almost two orders of magnitude deeper than for typical hot Jupiters. This means that the $P$-$T$ model should be sufficiently flexible at high pressures as the spectrum will be able to constrain it there. Indeed, we find that the slope of the $P$-$T$ profile at high pressures has a very strong effect on the high-flux regions of the spectrum.

A potential $P$-$T$ parametrisation is one with a series of free temperature parameters, which allows for considerable freedom in the retrieval. However, this can result in overfitting due to unphysical oscillations in temperature, which must be corrected for \citep[e.g.][]{Line2015,Kitzmann2020}. Our new $P$-$T$ model overcomes this issue by excluding certain unphysical, over-fitted $P$-$T$ profiles. The profile is characterised by the slopes of the $P$-$T$ profile across given atmospheric layers, and the priors on these slopes are chosen such that atmospheric temperature always increases with pressure. This eliminates large oscillations and makes the model more robust to changes in the number of $P$-$T$ parameters compared to a model with free temperature parameters. Although it would be preferable to allow for the possibility of finding a thermal inversion (noting that brown dwarfs are not expected to exhibit these), it would be difficult to distinguish the discovery of an inversion from over-fitting.

We then apply the adapted retrieval framework to the HST/WFC3 spectrum of the T5 brown dwarf 2MASS 2339+1352 \citep{Buenzli2014}. To test the accuracy of the method, we first apply it to a simulated spectrum with known input parameters. The retrieval is able to accurately retrieve these inputs, confirming that the method is self-consistent. When applied to 2MASS 2339+1352, the retrieval strongly constrains the mixing fractions of H$_2$O and CH$_4$ at $10^{-3.7}$ and $10^{-4.3}$, respectively, with 1$\sigma$ uncertainties of $\sim$0.1 dex (table \ref{tab:real_data}). These abundances suggest sub-solar elemental abundances for oxygen and carbon: $0.13 \pm 0.03$ and $0.07^{+0.03}_{-0.02}$ times solar values, respectively \citep{Asplund2009}. We place an upper limit of $10^{-5.9}$ on the mixing ratio of NH$_3$ (at the 99\% confidence level), which is expected from the retrieved $P$-$T$ profile as nitrogen is expected to be locked in N$_2$ at higher temperatures (e.g. above 500K at 1bar \citep{Burrows1999, Moses2013}).  We also place an upper limit of $10^{-6.9}$ on the mixing ratio of K and confirm that, in this spectral range, uncertainties in the Na and K cross sections do not have a significant effect since a retrieval which does not include these species produces the same results for all other model parameters. The retrieval also strongly constrains the $P$-$T$ profile in the photosphere between $\sim$0.6-20 bar. We estimate the effective temperature of 2MASS~J2339+1352 by extending the median retrieved model to longer wavelengths, obtaining a value of $\sim$1100~K. This is consistent with the expected range of effective temperatures for T5 dwarfs, i.e. $\sim$1000-1500~K \citep[e.g.][]{Kirkpatrick2005}.

Using a distance estimate for 2MASS~J2339+1352 \citep{Dupuy2012,Buenzli2014}, we further estimate its mass and radius based on our retrieval results. The values we obtain are 5$^{+3}_{-2}$ $M_\mathrm{J}$ and 1.1$\pm$0.2 $R_\mathrm{J}$ for the mass and radius, respectively. This radius estimate is in good agreement with the expected radius for a fairly young brown dwarf or giant planet \citep[e.g.][]{Baraffe2003}. The mass estimate, however, suggests that 2MASS~J2339+1352 could be a planetary object rather than a brown dwarf. We discuss two possible scenarios which can explain this. First, that 2MASS~J2339+1352 may potentially be a planetary mass object. Second, that the low mass and metallicity that we retrieve for 2MASS~J2339+1352 may be a result of the log($g$)-abundance degeneracy discussed in section \ref{sec:self-consistency}, and could potentially be rectified with a wider spectral coverage.

Microlensing observations have shown that unbound planets are very common \citep[e.g.][]{Sumi2011}. While the low metallicity we derive for this object could be consistent with an old age (and therefore inconsistent with a planetary mass, given the spectral type), this could also be a result of planetary formation processes. For example, \citet{Madhusudhan2014} find that formation through gravitational instability at large orbital separations can result in both sub-solar O and C abundances in giant exoplanets \citep[also see e.g.,][]{Helled2010,Oberg2011}. We also compare our derived mass and gravity estimates to evolutionary models in the literature. Assuming an effective temperature in the range 1000-1500~K, figure 9 of \citet{Burrows1997} suggests an age of $\sim 10^{6.5}-10^{7.5}$~yr, assuming a solar composition. \citet{Baraffe2003} predict a similar age ($\sim 10^7$~yr) for a 5M$_\mathrm{J}$ planet with an effective temperature in the range $\sim$1000-1500~K. For a sub-solar composition, this age estimate would be even lower \citep[e.g.][]{Burrows2001}. \citet{Sumi2011} find that unbound planetary-mass objects have a different mass function to field brown dwarfs, suggesting different formation pathways. In particular, they suggest that the planetary-mass objects may form in protoplanetary discs and subsequently be ejected \citep[e.g.][]{Veras2009}. If 2MASS~J2339+1352 is indeed a planetary-mass object, this would mean that it formed and was ejected from its stellar system very quickly. Since giant planet formation due to disk instability can occur on relatively fast timescales compared to core accretion (e.g. $\lesssim 10^4$~yr; \citealt{Durisen2007} and references therein), this may be possible in some scenarios \citep{Veras2009}.

An alternative explanation for the low mass and metallicity derived from the retrieval is that these quantities may be affected by the strong log($g$)-abundance degeneracy. As discussed in section \ref{sec:self-consistency}, both gravity and the abundances of the dominant species shaping the spectrum have similar effects on the emergent spectrum, such that a low gravity (i.e. low mass), low metallicity solution is similar to a higher gravity, higher metallicity solution. One way to break this degeneracy would be to use a wider spectral coverage, probing a larger number of molecular and/or atomic features. Other studies of brown dwarf and poorly-irradiated planet retrievals have also found unexpected and/or unphysical results for some parameters, in some cases possibly due to degeneracies between various model parameters. For example, \citet{Lavie2017} and \citet{Kitzmann2020} both find smaller radii than expected when an uninformative prior is assumed. Conversely, \citet{Todorov2016} find that gravity is not constrained in their retrieval of $\kappa$~Andromedae~b. This highlights the challenging nature of retrievals for isolated objects, for which the mass and radius are unknown and add further degeneracies compared to atmospheric retrievals of transiting exoplanets for which masses and radii are known a priori. However, future facilities such as JWST will have the capability of providing high-precision spectra over a wide spectral range, mitigating such biases and allowing for high-precision abundance estimates.

Our technique shows great potential for high-precision abundance determinations; the 0.1 dex precision we obtain here for the H$_2$O and CH$_4$ mixing ratios is comparable to the precision achieved with brown dwarf retrievals in the literature \citep[e.g.][]{Line2017,Zalesky2019}, and is significantly more stringent than the $\sim$ 0.5 dex achievable with current exoplanet emission spectra \citep[e.g.][]{Gandhi2018}. Given the availability of numerous high-quality brown dwarf spectra, this method can therefore enable accurate metallicity and compositional estimates for a large number of objects, allowing for detailed population studies and the testing of formation scenarios. Furthermore, given distance estimates, the retrieved radius-distance ratio and gravity can be used to place independent constraints on the masses and radii of brown dwarfs. Combined with self-consistent forward models, these estimates can contribute to the understanding of the physical processes and evolution of brown dwarfs. In the future, exoplanet spectra will also reach this level of quality, with large facilities such as the James Webb Space Telescope and the Extremely Large Telescope. The considerations investigated here therefore provide a step towards the accurate interpretation of these much-awaited observations.

This new retrieval framework for brown dwarfs provides insight for two areas of atmospheric characterisation: precise atmospheric characterisation of brown dwarfs and considerations for the analysis of high-quality exoplanet spectra in the future. As we have shown in this work, it is essential to understand the uncertainties of our models and their effects on data interpretation as we approach a new era of high-SNR exoplanet and brown dwarf spectroscopy.

\section*{Acknowledgements}
We would like to thank Siddharth Gandhi for sharing the HyDRA code and for useful discussions. We thank Daniel Apai for sharing data from his published work and for useful discussions. We also thank Nicole Allard for sharing her Na and K line profiles. A.A.A.P. acknowledges financial support from the Science and Technology Facilities Council (STFC), UK, towards her doctoral programme.

\section*{Data availability}
No new data were generated or analysed in support of this research.




\bibliographystyle{mnras}
\bibliography{refs} 



\appendix

\section{Retrieval without N\lowercase{a} and K}

\begin{figure*}
\centering
    \begin{subfigure}[b]{0.48\textwidth}
  		\includegraphics[width=\textwidth]{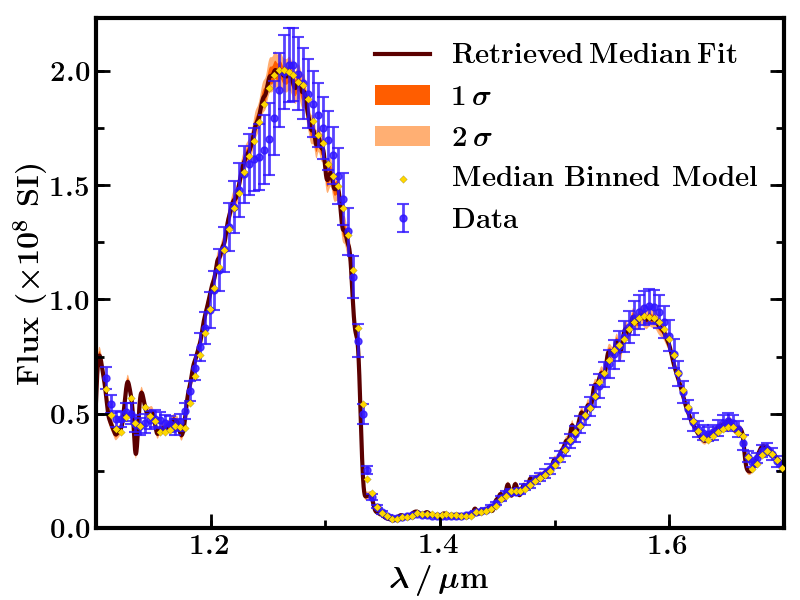}
    \end{subfigure}
    \begin{subfigure}[b]{0.48\textwidth}
    	\includegraphics[width=\textwidth]{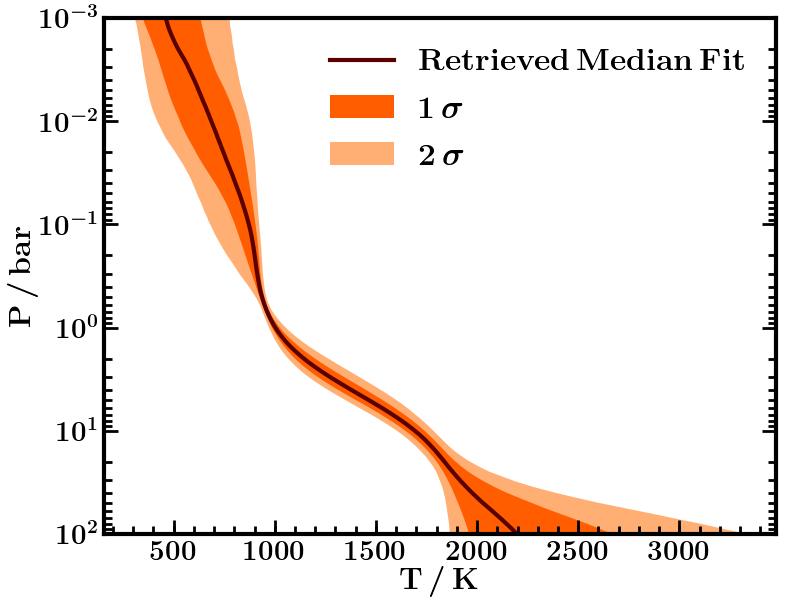}
    \end{subfigure}
\caption{\label{fig:nonak_spec} Spectrum and $P$-$T$ profile retrieved for 2MASS J2339+1352 without including Na and K in the retrieval model. Left: The median retrieved spectrum (maroon line) and 1$\sigma$ and 2$\sigma$ confidence intervals (dark and light orange shading, respectively). Data points are shown by blue circles, and error bars depict the median retrieved tolerance. Yellow diamonds show the median binned model. Flux axis is shown in SI units (Wm$^{-2}$m$^{-1}$). Right: Median retrieved $P$-$T$ profile (maroon line) and 1$\sigma$ and 2$\sigma$ confidence intervals (dark and light orange shading, respectively).}
\end{figure*}

\begin{figure*}
\centering
    	\includegraphics[width=0.9\textwidth]{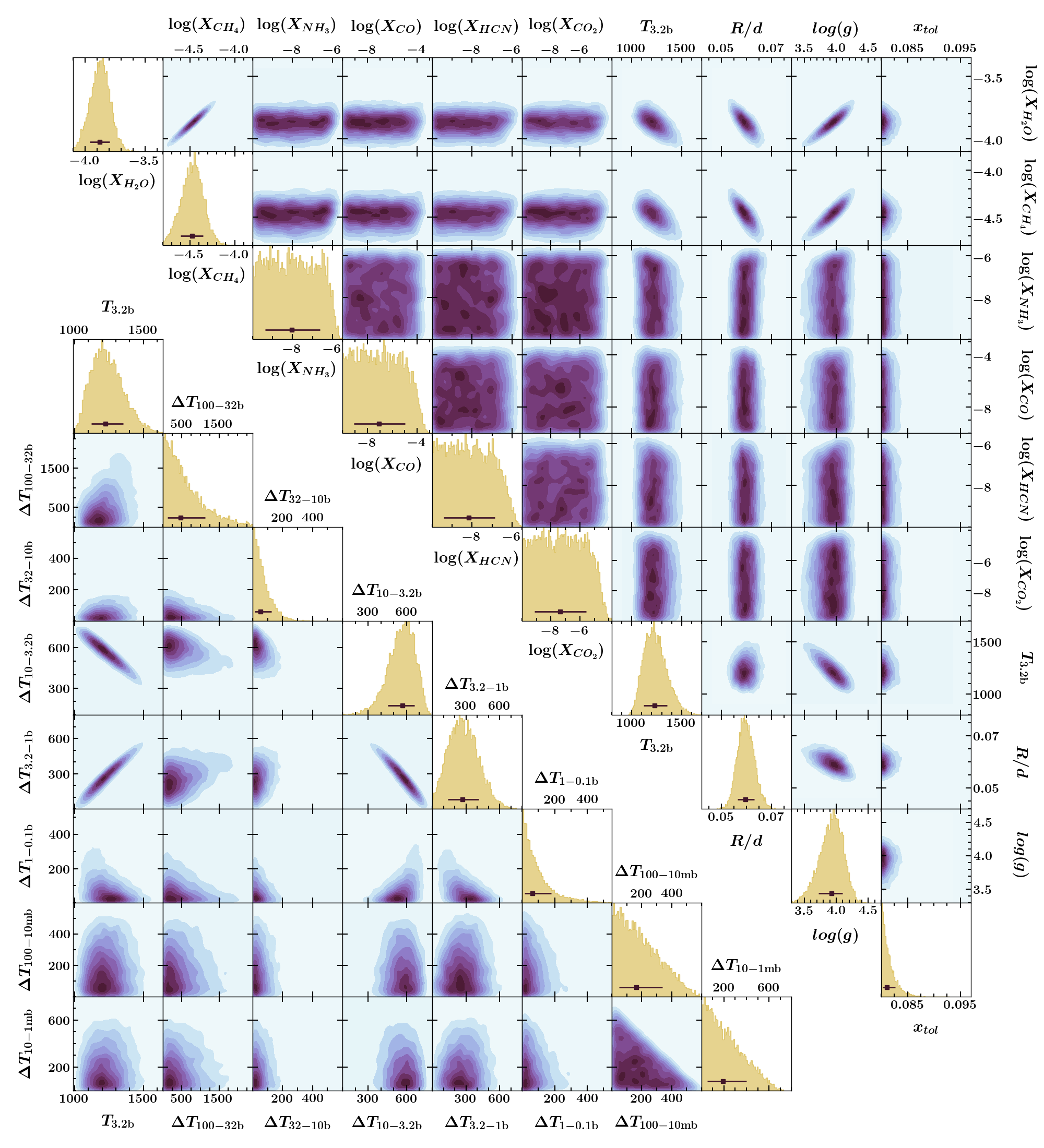}
\caption{\label{fig:nonak_corner} Marginalised posterior probability distribution from the retrieval of 2MASS J2339+1352 without Na and K. 1D marginalised posteriors are shown for each parameter along the diagonals, with median values and 68\% confidence intervals shown by the dark squares and error bars, respectively.}
\end{figure*}

In figures \ref{fig:nonak_spec} and \ref{fig:nonak_corner}, we show the results of the retrieval of 2MASS J2339+1352 without Na and K. The results are consistent with the retrieval which includes Na and K, as discussed in section \ref{sec:results}. However, when K is not included, the spectral feature at $\sim$1.25 $\mu$m is not fitted. The retrieved model parameter values are shown in table \ref{tab:nonak}

\begin{table}
\begin{centering}
\begin{tabular}{c|c|c}
Parameter &  Retrieved value  \\[3pt]
 \hline
$\textrm{log}(X_{H_2O})$ & $-3.88 \pm 0.08$  \\[7pt]
$\textrm{log}(X_{CH_4})$ & $-4.5 \pm 0.1$  \\[7pt]
$\textrm{log}(X_{NH_3})$ & $<-5.9$  \\[7pt]
$\textrm{log}(X_{CO})$ & $<-3.7$  \\[7pt]
$T_{3.2\textrm{b}}$ / K & $1230^{+130}_{-110}$  \\[7pt]
$R/d$  & $0.059^{+0.004}_{-0.003}$  \\[7pt]
$\textrm{log}(g)$ & $3.9 \pm 0.2$  \\[7pt]
$x_{tol}$  & $ 8.1^{+0.2}_{-0.08} \%$  \\[7pt]
\end{tabular}
\caption{\label{tab:nonak} A summary of the retrieved atmospheric properties of 2MASS J2339+1352 without including Na and K in the retrieval. These values are consistent with the results of the retrieval which includes Na and K (section \ref{sec:results})}
\end{centering}
\end{table}

\bsp	
\label{lastpage}
\end{document}